\documentclass[12pt]{iopart}
\usepackage{epsfig}

\newcommand{\bitem}{\begin{itemize}}
\newcommand{\eitem}{\end{itemize}}

\newcommand{\be}{\begin{equation}}
\newcommand{\ee}{\end{equation}}

\newcommand{\bea}{\begin{eqnarray}}
\newcommand{\eea}{\end{eqnarray}}
\newcommand{\beastar}{\begin{eqnarray*}}
\newcommand{\eeastar}{\end{eqnarray*}}

\newcommand{\lav}{\left\langle}
\newcommand{\rav}{\right\rangle}

\newcommand{\half}{\frac{1}{2}}

\newcommand{\eq}[1]{~(\ref{#1})}
\newcommand{\eqq}[2]{~(\ref{#1},\ref{#2})}

\newcommand{\order}{{{\mathcal O}}}

\newcommand{\ie}{{\it i.e.}}
\newcommand{\eg}{{\it e.g.}}

\newcommand{\rh}{\rho}
\newcommand{\sig}{\sigma}
\newcommand{\rhosig}{\rh(\sig)}
\newcommand{\rhoL}{\rh(L)}
\newcommand{\rhonsig}{\rh\pn(\sig)}

\newcommand{\rhoalsig}{\rh\pa(\sig)}

\newcommand{\musig}{\mu(\sig)}
\newcommand{\intsig}{\int\!d\sig\,}
\newcommand{\intL}{\int\!dL\,}
\newcommand{\w}{w}
\newcommand{\wi}{w_i(\sig)}

\newcommand{\ri}{\rh_i}
\newcommand{\rn}{\rh_0}
\newcommand{\ro}{\rh_1}
\newcommand{\rt}{\rh_2}

\newcommand{\lam}{\lambda}

\newcommand{\lamo}{\lam_1}
\newcommand{\lami}{\lam_i}

\newcommand{\fexc}{\tilde{f}}

\newcommand{\muexci}{\tilde{\mu}_i}
\newcommand{\muexc}{\tilde{\mu}}

\newcommand{\fmom}{f_{\rm mom}}

\newcommand{\al}{\alpha}
\newcommand{\ph}{v}

\newcommand{\pn}{^{(0)}}
\newcommand{\po}{^{(1)}}
\newcommand{\pa}{^{(\al)}}
\newcommand{\pb}{^{(\beta)}}
\newcommand{\ptwo}{^{(2)}}

\newcommand{\prob}{n}
\newcommand{\probsig}{\prob(\sig)}
\newcommand{\probnsig}{\prob\pn(\sig)}

\newcommand{\mom}{moment density}
\newcommand{\moms}{moment densities}

\newcommand{\eps}{\epsilon}
\newcommand{\lmax}{L_{\rm max}}
\newcommand{\ls}{L_{\rm s}}

\newcommand{\sav}{\bar{\sig}}
\newcommand{\st}{s_{\rm t}}
\newcommand{\rv}{{\bf r}}

\newcommand{\Ln}{\tilde{L}}
\newcommand{\Dn}{\tilde{D}}

\begin{document}

\topical{Predicting phase equilibria in polydisperse systems}

\author{Peter Sollich}

\address{Department of Mathematics, King's College London, Strand,
London WC2R 2LS, U.K. Email: {\tt peter.sollich@kcl.ac.uk}}

\begin{abstract}
Many materials containing colloids or polymers are polydisperse: They
comprise particles with properties (such as particle diameter, charge,
or polymer chain length) that depend continuously on one or several
parameters. This review focusses on the theoretical prediction of
phase equilibria in polydisperse systems; the presence of an
effectively infinite number of distinguishable particle species makes
this a highly nontrivial task. I first describe qualitatively some of
the novel features of polydisperse phase behaviour, and outline a
theoretical framework within which they can be explored. Current
techniques for predicting polydisperse phase equilibria are then
reviewed. I also discuss applications to some simple model systems
including homopolymers and random copolymers, spherical colloids and
colloid-polymer mixtures, and liquid crystals formed from rod- and
plate-like colloidal particles; the results surveyed give an idea of
the rich phenomenology of polydisperse phase behaviour.  Extensions to
the study of polydispersity effects on interfacial behaviour and phase
separation kinetics are outlined briefly.
\end{abstract}


\submitto{\JPCM}


\section{Introduction and scope}

Statistical mechanics was originally developed for the study of large
systems of identical particles such as atoms and small
molecules. However, many materials of industrial and commercial
importance which contain colloidal particles or polymers do not fit
neatly into this framework. For example, the particles in a colloidal
suspension are never precisely identical to each other, but have a
range of radii (and possibly surface charges, shapes etc).
Industrially produced polymers always contain macromolecules with a
range of chain lengths; and hydrocarbon mixtures occurring in the
petrochemical industry often consist of a large number of different
molecular species best described as having continuously varying
properties across each family of molecules. All these materials are
therefore {\em polydisperse}: They contain particles with properties
depending continuously on one or several parameters.

In this review, I will focus on the effects of polydispersity on {\em
phase behaviour}: To process a colloidal or polymeric material, one
needs to know under which conditions of pressure and temperature it
will be stable against demixing, how many phases will result if it
does demix, and what their properties are. The emphasis will be on the
problem of predicting such phase behaviour theoretically, although
I will complement this by references to experimental observations and the
results of computer simulations where appropriate. I will concentrate
almost exclusively on {\em bulk phase equilibria}, giving only the
briefest outlook towards the treatment of inhomogeneous systems
(interfacial behaviour etc) and the challenging topic of phase separation
kinetics in Sec.~\ref{sec:outlook}. Finally, I will only discuss the
case of {\em fixed polydispersity}, where the polydisperse attribute
of each particle remains fixed once and for all. This includes all the
examples given above; the length of a polymer molecule or the
size of a colloidal particle, for example,
do not change over time.  The contrasting
case of variable polydispersity is exemplified by a surfactant
solution in which the surfactant molecules form worm-like micelles
whose lengths constantly change due to scission and
recombination~\cite{CatCan90}. Systems of this kind have been
treated theoretically (see \eg~\cite{StaTilSluQui88,StaTilQui90}) but
will be excluded below because their phase behaviour is much less
complex than that of fixed polydispersity systems; the reasons for
this are explained in Sec.~\ref{sec:basics} below.

Below, I will first explain why polydisperse phase equilibria are
challenging to predict, describe some of the new effects that can
occur, and outline a theoretical framework within which they can be explored
(Sec.~\ref{sec:basics}). Then I will give an overview of some current
techniques for predicting polydisperse phase equilibria
(Sec.~\ref{sec:methods}). Applications to some simple model systems
are discussed in Sec.~\ref{sec:systems}, with the aim of giving an
idea of the rich phenomenology of polydisperse systems.
Sec.~\ref{sec:outlook} describes briefly the considerable challenges
that one faces when looking at polydispersity effects on interfacial
behaviour and phase separation kinetics.

Because of the volume of the literature, I will not attempt to give a
historical account of the development of theoretical work on
polydisperse phase behaviour. The following, very selective, sketch
will have to suffice: De Donder's work in the 1920's on many-component
mixtures~\cite{Donder27} is often cited as an early and important
precursor. From the 1940's onwards, there were significant
contributions in the area of polydisperse (homo- and co-)polymers,
associated with the names of Flory, Huggins, Koningsveld, Scott,
Solc and Stavermann~\cite{Flory44,ScoMag45,Scott45,Scott52,%
KonSta67,KonSta67b,KonSta68,Solc70} among many others. Around the
same time, the concept of polydispersity also appeared in the
treatment of the distillation of multi-component hydrocarbon mixtures
(see \eg~\cite{Bowman49,Hoffman68}). Since then polydispersity has
been recognised as important in many other contexts, notably the phase
behaviour of suspensions of spherical~\cite{BluSte79,Dickinson80} and
non-spherical (\eg\ rod-like~\cite{FloAbe78}) colloidal particles.

\section{Polydisperse phase equilibria}
\label{sec:basics}

\subsection{The challenge}

To understand why the prediction of phase equilibria in polydisperse
systems is a challenging problem, it is useful to recall first the
procedure for a monodisperse system. In a suspension of
identical colloidal particles, for example, the experimentally
controlled variables would be the temperature $T$, the suspension
volume $V$, and the number $N$ of colloidal particles; the appropriate
thermodynamic ensemble is therefore the canonical one, and the
thermodynamic potential is the Helmholtz free energy $F(N,V,T)$.  Here
I assume (and will do so throughout in what follows) that the solvent
degrees of freedom have been formally eliminated, so that $F$
includes the effects of the solvent only through any effective
interaction it may mediate between the colloids. The suspension will
separate into two phases with particle numbers $N\pa$ and volumes
$V\pa$ ($\al=1,2$) if it can thereby lower its total free energy
$\sum_\al F(N\pa,V\pa,T)$ below the value $F(N,V,T)$. The $N\pa$
and $V\pa$ adopt the values which minimize this total free energy,
subject to conservation of volume ($\sum_\al
V\pa=V$) and particle number ($\sum_\al
N\pa=N$). Introducing Lagrange multipliers for these constraints then
gives the familiar coexistence conditions of equal chemical potential
$\mu=\partial F/\partial N$ and pressure $\Pi=-\partial F/\partial V$
in the two phases. (Because of the elimination of the solvent degrees
of freedom, $\Pi$ is actually the osmotic pressure of the colloids,
rather than the total suspension pressure.)

\begin{figure}
\begin{center}
\epsfig{file=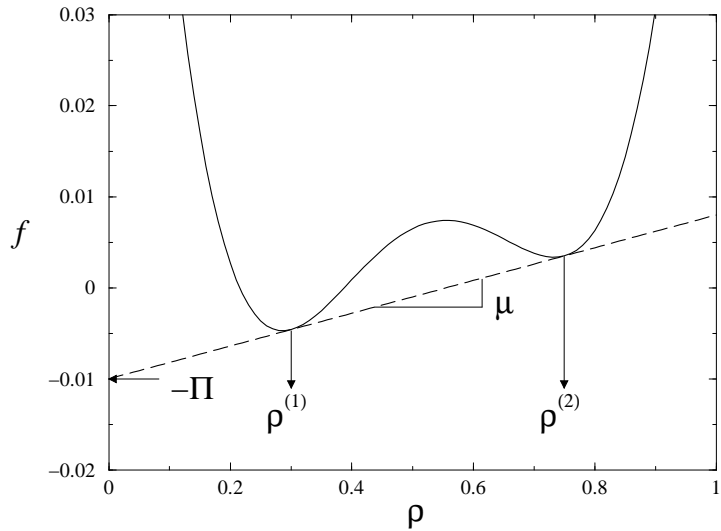,width=0.6\textwidth}
\end{center}
\caption{The double tangent construction for finding phase
equilibria (for monodisperse systems). Shown is a sketch of a free
energy density $f$ versus particle density $\rh$ (solid line) and the
double tangent to it (dashed line). The tangency points identify the
densities $\rh\po$ and $\rh\ptwo$ of the coexisting phases. The slope
and the negative intercept of the tangent give the chemical potential
$\mu$ and the osmotic pressure $\Pi$, respectively; both are common to
the coexisting phases. A parent phase with a density intermediate
between $\rh\po$
and $\rh\ptwo$ will phase separate into two phases with these
densities, thereby lowering the total free energy of the system.
\label{fig:double_tangent}
}
\end{figure}

Since the free energy is extensive, it can be written as
$F(N,V,T)=Vf(\rho,T)$ where $\rho=N/V$ is the (number) density of
colloids. In terms of the free energy {\em density} $f(\rho,T)$, the
coexistence condition has a simple geometrical interpretation. From
the definition of $f$, one has $\mu=\partial(Vf(N/V,T))/\partial N = 
\partial f/\partial \rho$ and
\be
\Pi= -\frac{\partial}{\partial V} (Vf(N/V,T)) =
-f + \rho\, \partial f/\partial \rho = -f+\mu\rho
\label{Pi_monodisp}
\ee
(the latter result can also be seen directly from the Gibbs-Duhem
relation $F+\Pi V -\mu N = 0$). Plotting $f$ as a function of $\rho$
as in Fig.~\ref{fig:double_tangent}, one sees that $\mu$ gives the
slope of the tangents to the plot at the densities of the two
coexisting phases, and that $-\Pi$ is their intercept with the
$f$-axis. Since $\mu$ and $\Pi$ are equal, so are the tangents
themselves: The
densities $\rh\po$ and $\rh\ptwo$ of the coexisting phases are
determined by constructing a {\em double tangent} to $f(\rho,T)$ (see
\eg~\cite{DeHoff92}). From these densities one can then find the
fraction of the system volume $\ph\pa=V\pa/V$ occupied by each phase, by
using particle conservation: Dividing $\sum_\al N\pa = N$ by $V$
one has $\sum_\al (V\pa/V)(N\pa/V\pa)=N/V$ or $\sum_\al \ph\pa
\rh\pa = \rho$; for two phases, using $\ph\po+\ph\ptwo=1$, this gives
the well-known ``lever rule'' $\ph\po =
(\rh-\rh\ptwo)/(\rh\po-\rh\ptwo)$.

Moving towards the polydisperse case, assume now that there are $M$
different species of colloid particles, each with its own particle
number $N_i$ and corresponding density $\rho_i=N_i/V$. All densities
are conserved, so that
\be
\sum_\al \ph\pa \rho\pa_i = \rh_i
\label{discrete_cons}
\ee
if the system separates into several phases. The free energy density
$f(\{\rh_i\})$ is now a function of all $M$ densities, as well as the fixed
temperature $T$ which I suppress from now on in the notation. (I will
also call $f$ simply the free energy rather than the free energy
density where no misunderstanding is possible.) A plot of $f(\{\rh_i\})$
against the densities $\rh_i$ would give 
a (hyper-)surface in a graph with $M+1$
coordinate axes, and to find phase coexistences we would have to
construct multiple tangent (hyper-)planes to this surface.  Where such
tangent planes exist, the total free energy is lowered by phase
separation into the appropriate number of phases (which, from Gibbs'
phase rule, can be between two and $M+1$). The densities $\rho_i\pa$
in the different phases are given by the points where the tangent
plane touches the free energy surface, and the fractional phase volumes
$\ph\pa$ follow from the
conditions\eq{discrete_cons} together with $\sum_\al \ph\pa = 1$.

Now consider the fully polydisperse case. Let $\sig$ be the
polydisperse attribute of the particles, \eg\ the particle diameter in
spherical colloids or the chain length in polymers. To fully describe
the composition of the system we now need a {\em density distribution}
$\rhosig$, defined such that $\rhosig d\sig$ is the density of
particles with $\sig$-values in the range
$[\sig,\sig+d\sig]$. Formally, this corresponds to a scenario with an
infinite number of particle species, as can be seen by splitting the
range of $\sig$ into $M$ ``bins'', defining the $\rho_i$ to be the
densities within each bin, and then taking $M\to\infty$ (see
\eg~\cite{SalSte82}). The tangent plane procedure for finding phase
coexistences then clearly becomes unmanageable, both
conceptually and numerically: One would have to work in an
infinite-dimensional space---which mathematically corresponds to the
fact that the free energy becomes a {\em functional} $f([\rhosig])$ of
the density distribution $\rhosig$---and Gibbs' phase rule allows the
coexistence of arbitrarily many thermodynamic phases. 

In summary, then, the challenge in predicting polydisperse phase
equilibria arises from the effectively infinite number of conserved
densities. This renders the standard approaches developed for mixtures
with a finite number of species useless. Note that the difficulty that
I am talking about here is that of determining the phase equilibria
from a free energy (functional) which is assumed {\em known}. The
calculation of this free energy (or at least of a good approximation
to it) is a different---and no less challenging---problem that I will
not address in this review. So, in what follows, I will regard each
model free energy as given, and do {\em not} discuss in detail the
issue of how good a description of the real system it offers, nor how
or whether it can be derived from an underlying microscopic
Hamiltonian. Whenever I refer to ``exact'' results, I mean the exact
thermodynamics of such a model as specified by its free energy.

Finally, having established the presence of an infinite number of
conserved densities as the principal obstacle in the prediction of
polydisperse phase behaviour, one easily sees why {\em variable}
polydispersity is so much easier to deal with: There, one normally
fixes the ratios of the densities $\rho_i$ of the various species to
each other in the low density
limit~\cite{StaTilSluQui88,StaTilQui90}. (In the fully polydisperse
case this corresponds to fixing in the same limit the shape, but not
the overall scale, of the
density distribution $\rhosig$.) However, in this limit the densities
are directly related to the chemical potentials, and so one is
effectively fixing all chemical potential differences. The
thermodynamic variables are then $N$, $V$, $T$ and the chemical
potential differences, and so there is only a {\em single} conserved
density, just as in the monodisperse case. So, while the actual
determination of the relevant (semi-grandcanonical) free energy function might
still be a challenging problem, once this function is found the
determination of the phase behaviour can proceed by a standard double
tangent construction, and Gibbs' phase rule remains the same as for a
monodisperse system.

\subsection{Polydispersity gives rich phase behaviour}

\begin{figure}
\begin{center}
\epsfig{file=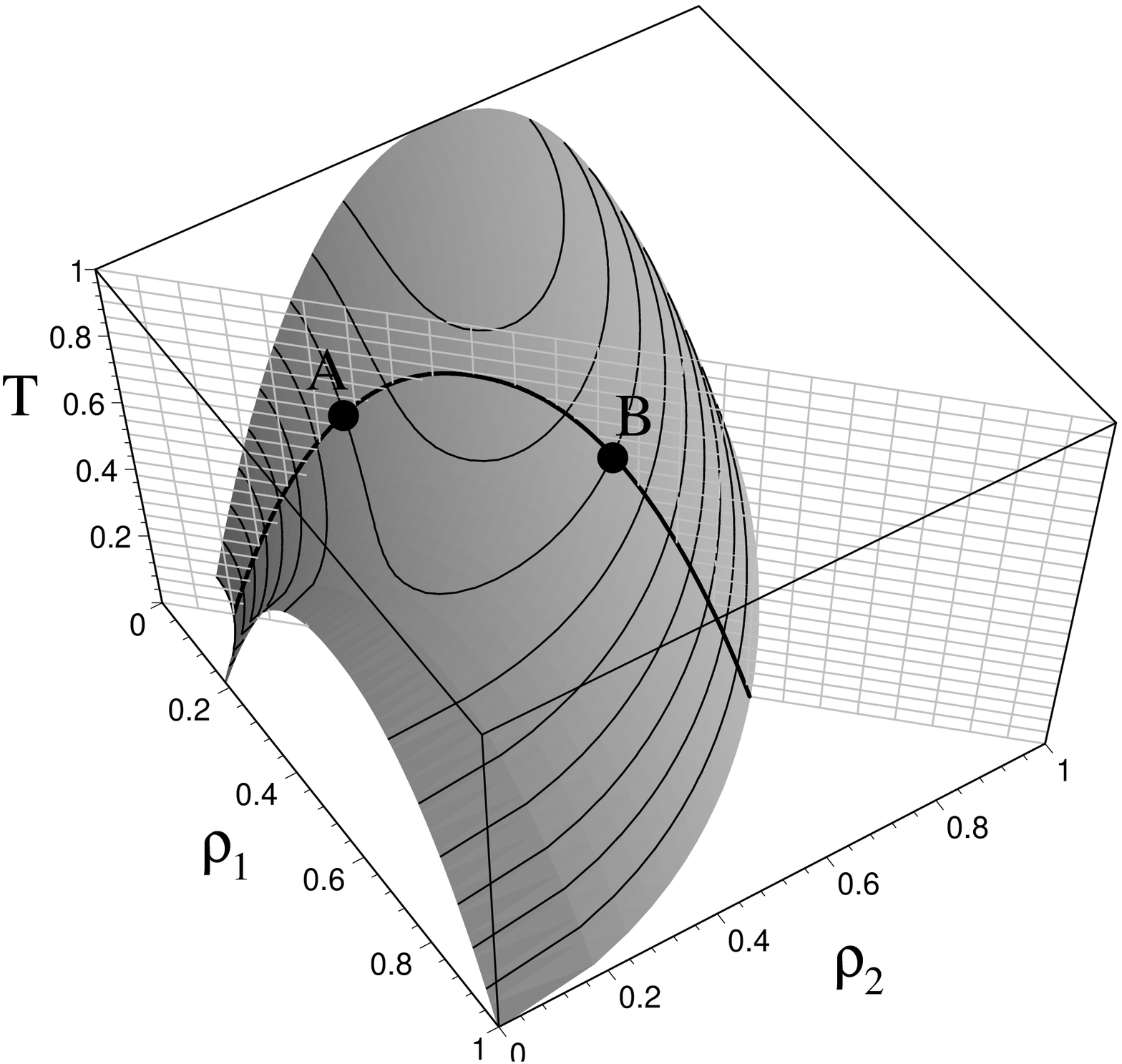,width=0.4\textwidth}

\vspace*{0.4cm}

\epsfig{file=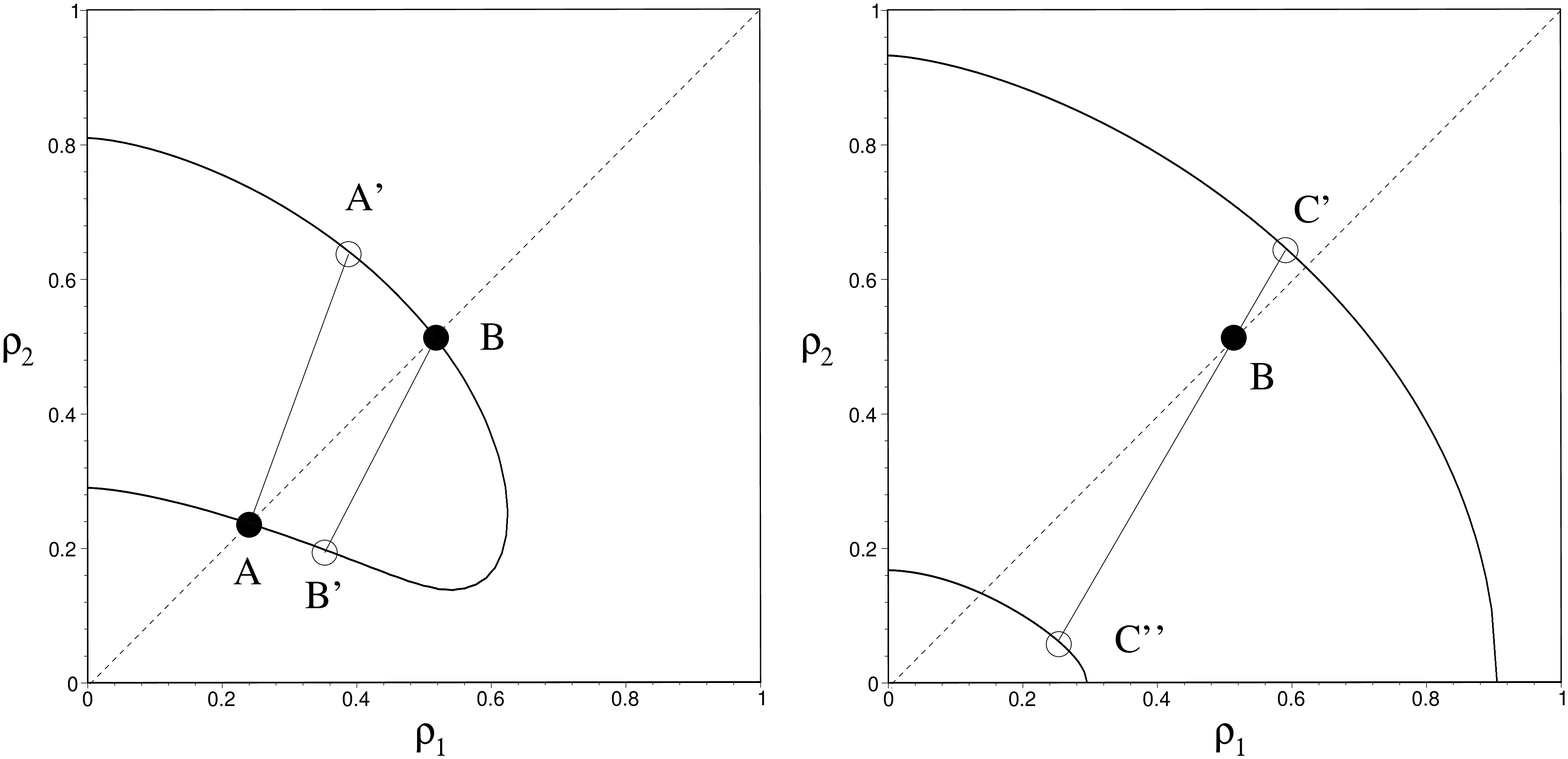,width=0.8\textwidth}
\end{center}

\caption{Top: A schematic phase diagram for a bidisperse system. The
surface shown in this $\ro$-$\rt$-$T$ plot delimits the region where
phase separation into two phases occurs. The vertical plane
corresponds to systems on a ``dilution line'', for which the
composition is fixed (here: same number of particles of species 1 and
species 2) but the overall density can vary. The intersection of this
plane with the phase boundary gives the cloud curve shown in
Fig.~\protect\ref{fig:CPC_demo}. Bottom left: Horizontal cut through
the phase diagram, corresponding to a fixed value $T_1$ of the
temperature. The dashed line is the dilution line. The filled circles,
marking the points where the dilution line intersects the phase
boundary, give the densities $(\ro,\rt)$ in the cloud phases, which
by definition begin to phase separate at the given temperature $T_1$. Tielines
connect the cloud phases with the coexisting shadow phases (empty
circles); these have different compositions from the cloud phases
since they are not located on the dilution line.  Bottom right: The
situation for a lower temperature $T_2$. Phase B, which at $T_1$ had
separated off an infinitesimal amount of B', has now separated into
two phases C' and C'' which are both present in nonzero amounts;
neither of them has the same composition as B.
\label{fig:three_d_phasediag}
}
\end{figure}

\begin{figure}
\begin{center}
\epsfig{file=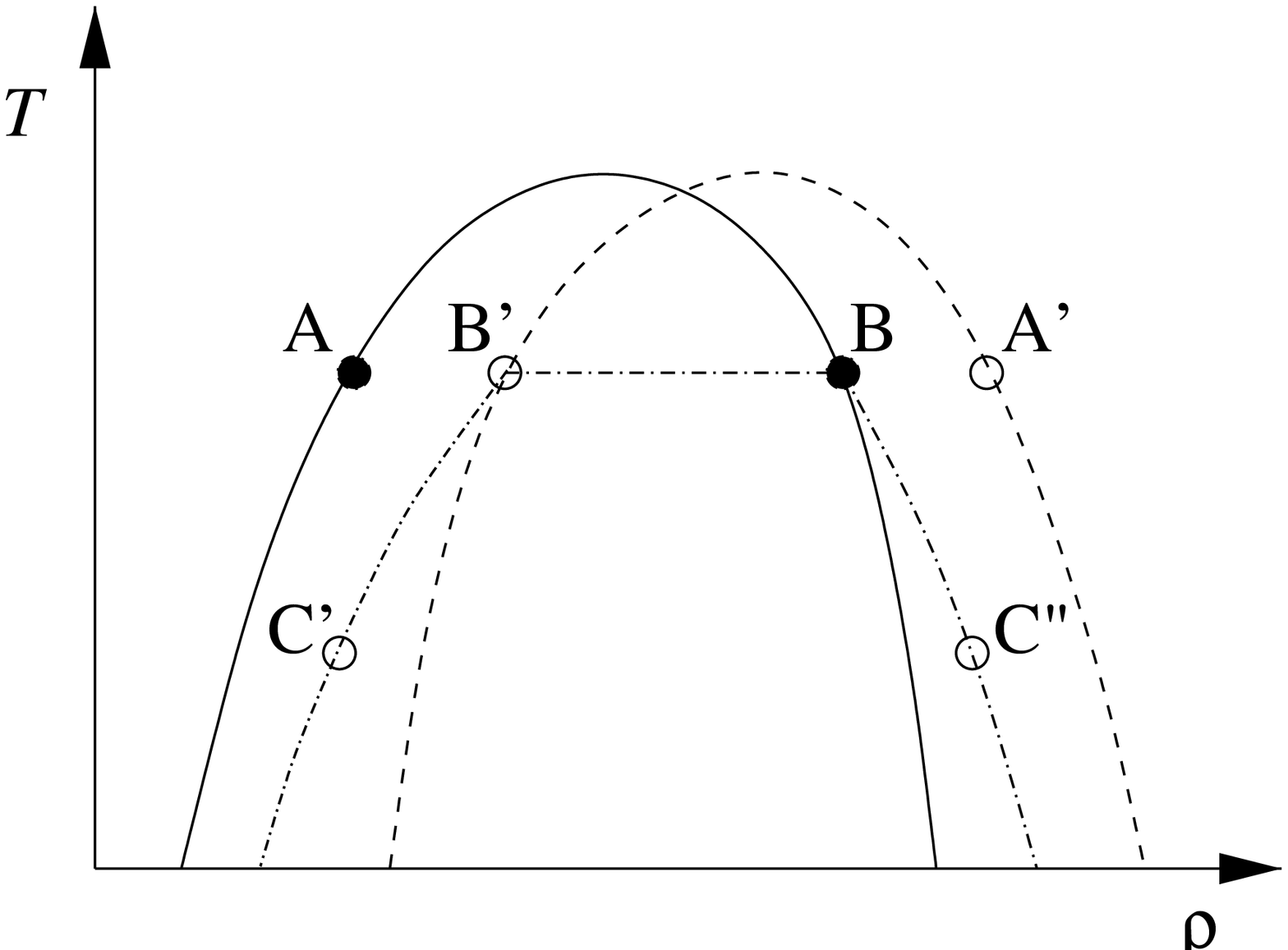,width=0.45\textwidth}%
\epsfig{file=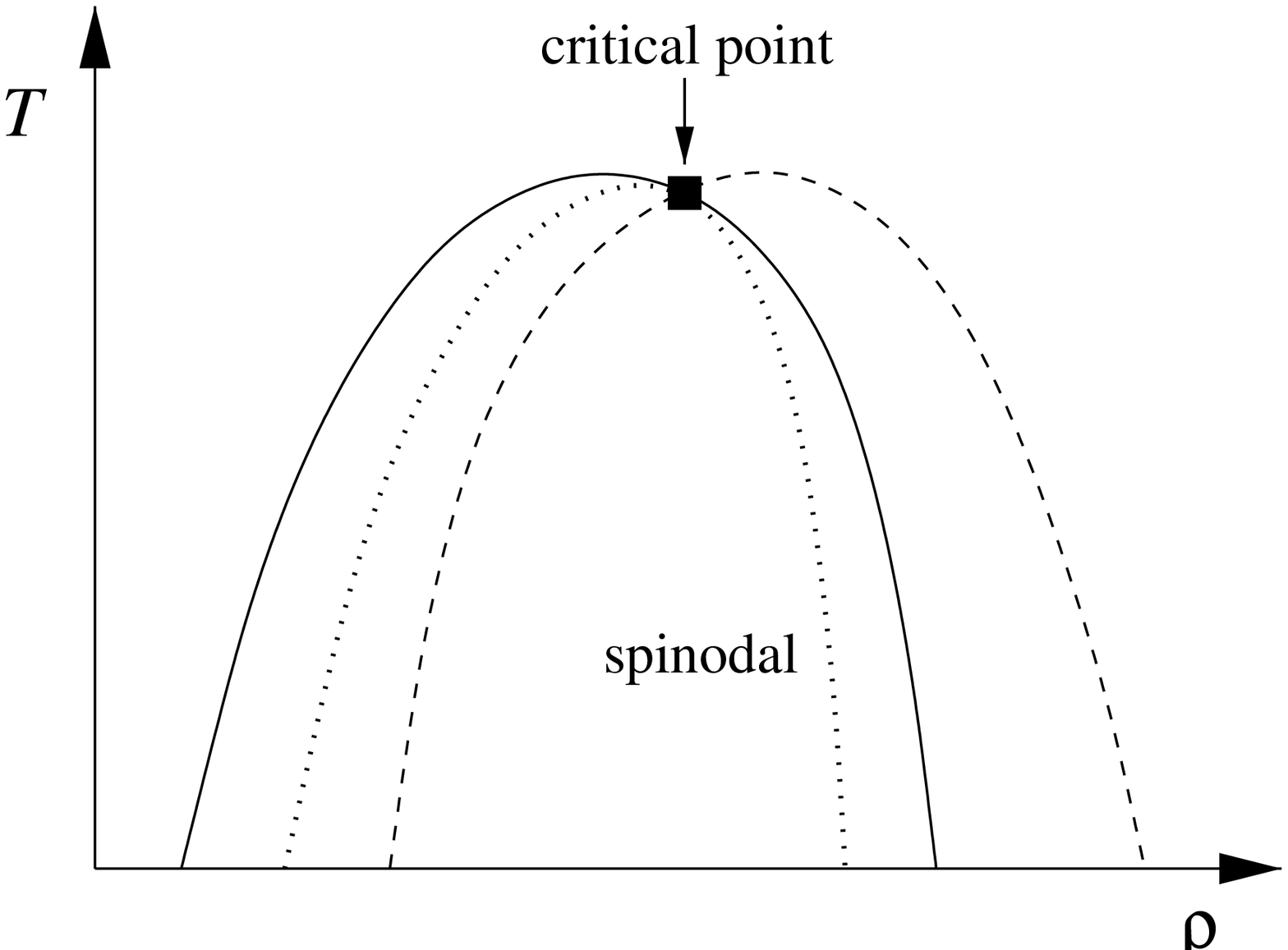,width=0.45\textwidth}
\end{center}
\caption{Left: Cloud (solid) and shadow (dashed) curves, schematically drawn
for the phase diagram of Fig.~\protect\ref{fig:three_d_phasediag}. On
the $x$-axis is the total particle number density $\rh=\ro+\rt$;
points corresponding to those in
Fig.~\protect\ref{fig:three_d_phasediag} are labelled with the same
letters. The dash-dotted lines are the coexistence curves for the
parent phase B, giving the densities of the phases
into which B separates as the temperature is decreased.
Right: Spinodal curve and critical point for the
same phase diagram. The critical point lies at an intersection of
cloud and shadow curves; the spinodal (dotted) lies inside the cloud
curve and touches it at the critical point.
\label{fig:CPC_demo}
}
\end{figure}

To try to understand the qualitative features of polydisperse phase
behaviour, it is useful to consider first a bidisperse system (with two
particle species), for which it is still possible to represent the full
$\ro$-$\rt$-$T$ phase diagram graphically. A schematic example of such
a phase diagram---inspired by the phenomenology of binary liquids---is
shown in Fig.~\ref{fig:three_d_phasediag}, with some tielines drawn
that connect coexisting phases. (A nice illustration based on the
Flory-Huggins theory of polymer solutions is given
in~\cite{Koningsveld69}.) Assume the overall densities of the two
particles species are $\ro\pn$ and $\rt\pn$; I use the ``(0)''
superscript here to distinguish the properties of this ``parent''
phase from other generic values of $\ro$ and $\rt$.  At high
temperatures, the given composition of the system will be stable as a
single phase. As $T$ is lowered, however, the system will eventually
become able to reduce its total free energy by separating into several
(in this case: two) phases. The first temperature where this happens
defines the so-called ``cloud point''. At this point, the parent
coexists with an infinitesimal amount of a new phase, called the
``shadow'' phase~\cite{KonSta67,KonSta67b}. One can repeat this
procedure of finding the onset of phase coexistence for a different
parent, obtained by diluting with additional solvent; this just
changes the total density $\ro\pn+\rt\pn$ but preserves the ratio of
the densities of the two species. Plotting the cloud point temperature
against the total parent (cloud phase) density and against the total
density of the shadow, for a series of such diluted parents, one
obtains the so-called cloud curve and shadow curve, respectively
(Fig.~\ref{fig:CPC_demo}). In a monodisperse system, these two curves
would coincide, with a critical point at the maximum. In the
bidisperse (and more generally the polydisperse) case, however, cloud
point and shadow curve are different, and the critical point occurs at
a crossing of the two curves.

To understand this difference between monodisperse and polydisperse
systems, it is useful to bear in mind that the set of parent phases
whose behaviour is represented by the cloud curve have values of $\ro$
and $\rt$ which lie on a line through the origin in the $\ro$-$\rt$
plane. At a given temperature, the cloud point phases are found as the
intersections of this ``dilution line'' with the boundaries of the
region where phase coexistence occurs, while the corresponding shadow
phases are given by the opposite ends of the tielines starting at the cloud
point phases (see Fig.~\ref{fig:three_d_phasediag}). In general, the
shadows therefore do not lie on the dilution line; compared to the
dilution line composition which all cloud phases share, the shadow
phases have become enriched in one or the other of the two species, a
process normally referred to as ``fractionation''. Thus, in contrast to a
monodisperse system, the roles of cloud and shadow phases cannot be
reversed, and cloud and shadow curves are therefore in general
different. The fact that the critical point is located at a crossing
of the cloud and shadow curves (rather than at their maximum) follows
because at criticality cloud and shadow are by definition identical.

It is also sometimes useful to consider spinodals in a polydisperse
system; these are the points where (as temperature is varied, for
example) a given parent phase first becomes unstable to local
density fluctuations. Determining the
spinodal points for all parents on a dilution line gives a spinodal
curve which can be plotted along with the cloud and shadow curves. By
construction, outside the cloud curve single phases
are stable against phase separation, so the spinodal curve must lie
inside the cloud curve; the critical point always lies on the spinodal
curve (since the shadow phase there can be generated by an
infinitesimally small fluctuation) and so the spinodal and the cloud curve
touch there (see Fig.~\ref{fig:CPC_demo}).

Beyond the onset of phase coexistence, polydisperse phase behaviour
becomes yet more complex. Continuing with our bidisperse example, a
given parent phase will start to phase separate at the cloud point as
$T$ is lowered. For lower temperatures, two phases will coexist in
finite amounts; at each given $T$, the densities $\ro$ and $\rt$ in
these phases can be found from the ends of the unique tieline (in the
$\ro$-$\rt$ plane) that passes through the parent. Neither of the
coexisting phases will therefore be on the dilution line (see
Fig~\ref{fig:three_d_phasediag}), and both will contain different
fractions of particles of the two species; only the overall
composition across the two phases will be maintained. Plotting the
temperature against the total density of the two phases would generate
two ``coexistence'' curves which begin on the cloud and shadow curve,
respectively (see Fig.~\ref{fig:CPC_demo}). Each parent on the
dilution line will generate its own set of coexistence curves, all
beginning at different points on the cloud and shadow curves.

So far I have only discussed situations where at most two phases
coexist once phases separation occurs. In a polydisperse system, this
need not be the case, of course; as discussed above, there is no a
priori limit on the number of coexisting phases. To see the
qualitative effect of this on the representations of phase behaviour
\begin{figure}
\begin{center}
\epsfig{file=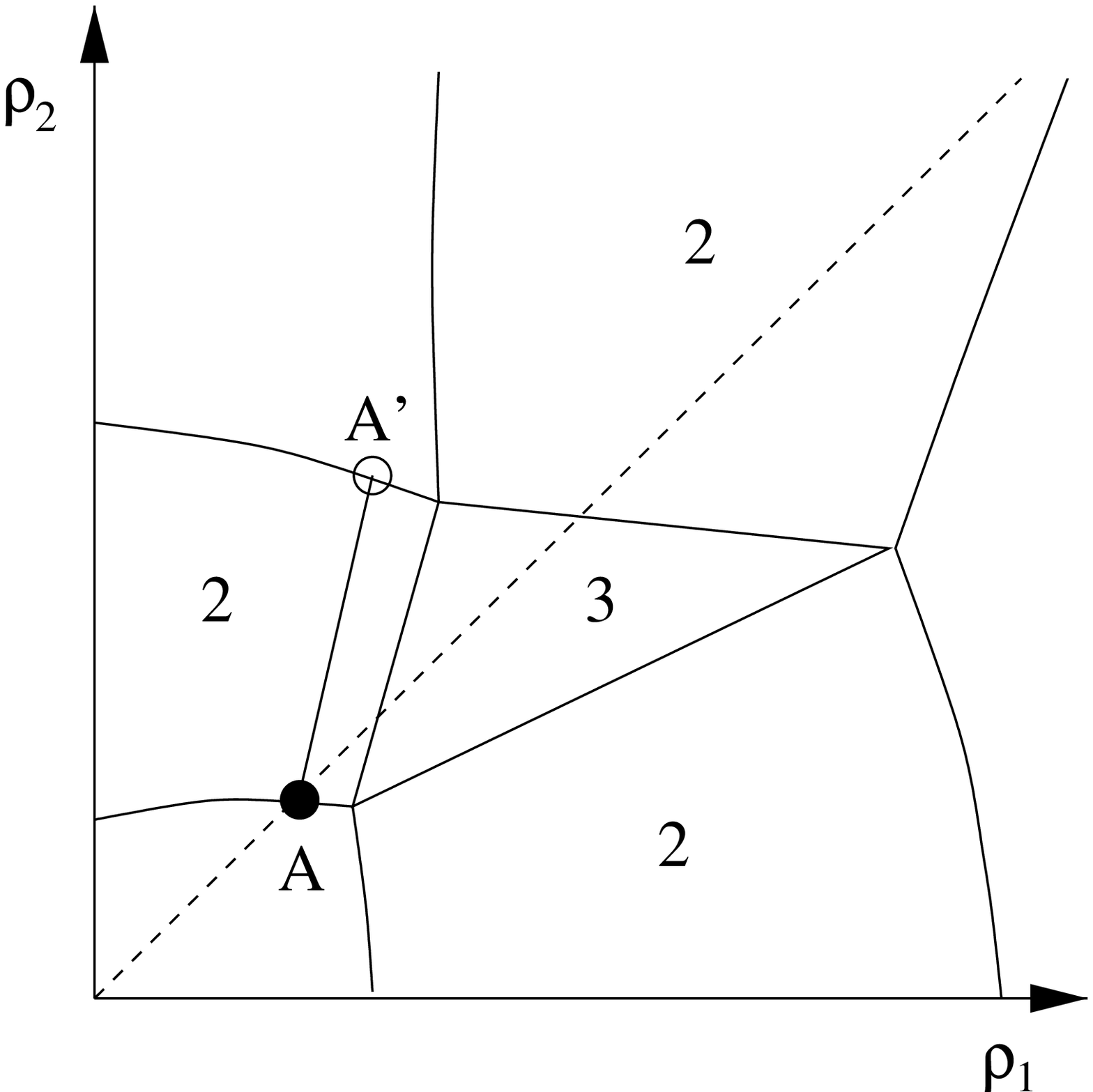,width=0.25\textwidth}\hspace*{3mm}
\epsfig{file=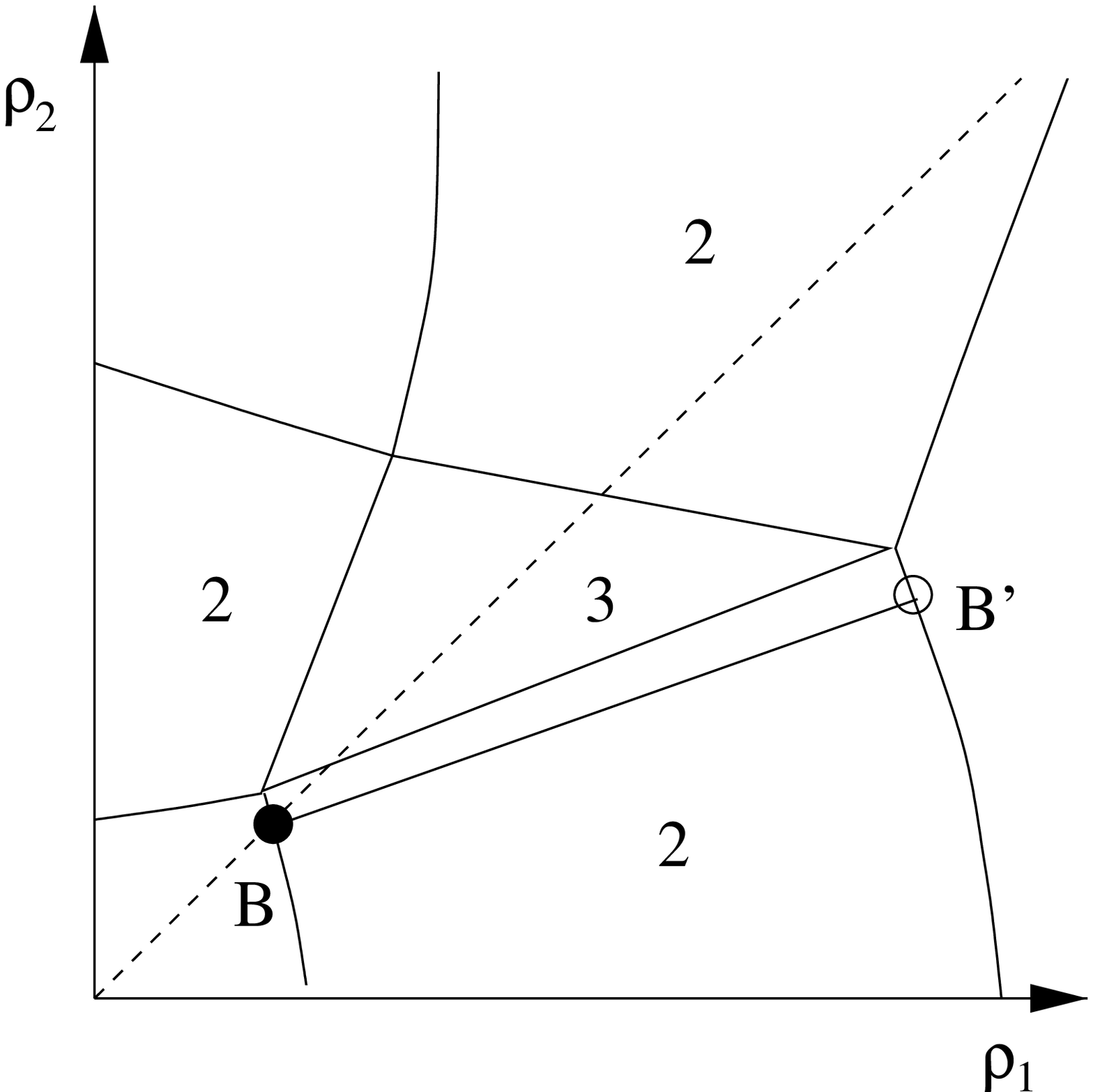,width=0.25\textwidth}\hspace*{3mm}
\epsfig{file=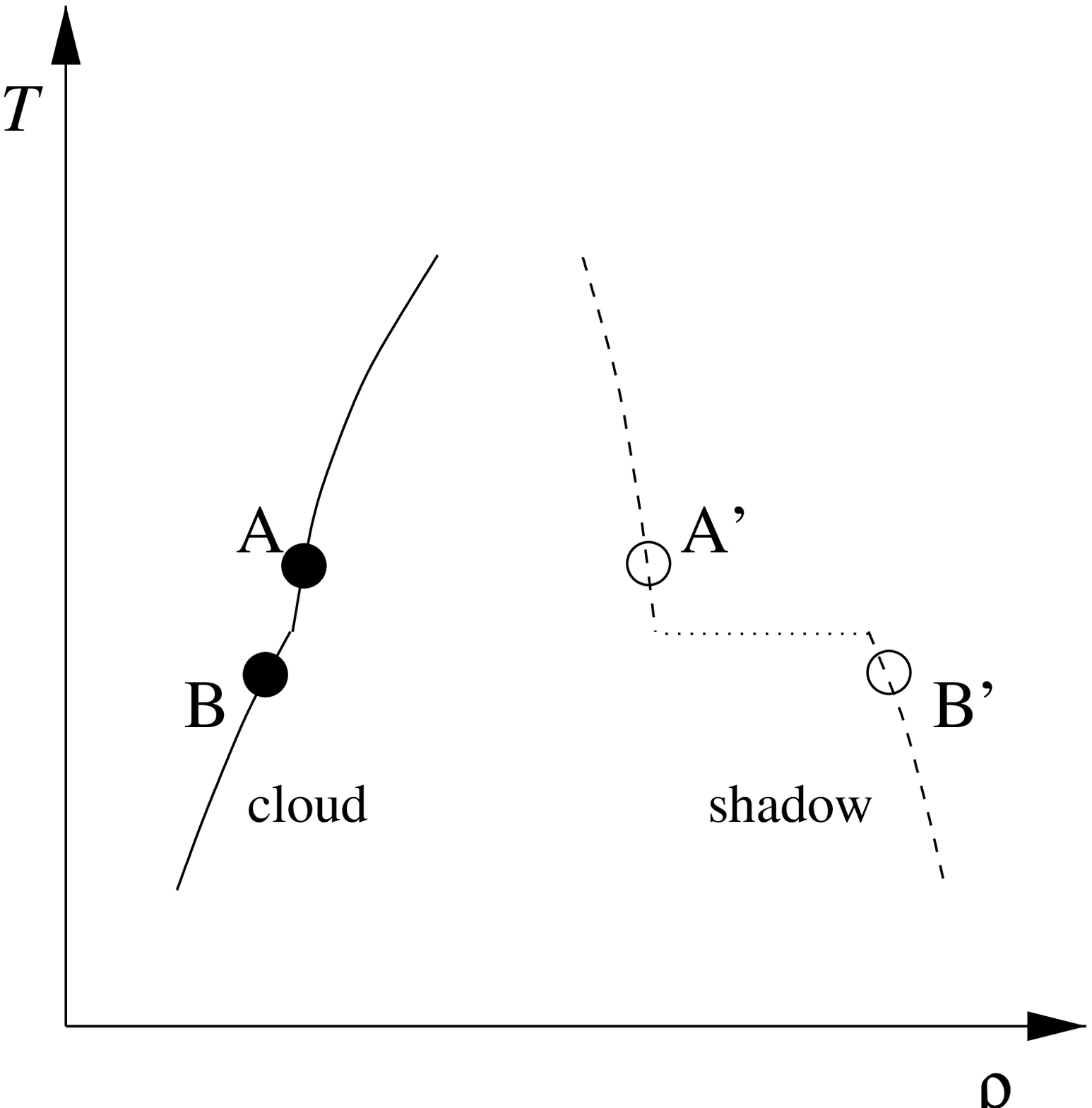,width=0.25\textwidth}
\end{center}
\caption{Left and middle: Constant temperature slices through the phase
diagram of a bidisperse system, with two- and three-phase regions as
indicated. On the left, the dilute cloud phase A (solid circle) begins
to phase separate by splitting off a denser phase A' containing
predominantly particles of species 2. On changing $T$ (middle plot),
the corner of the three-phase triangle may pass through the dilution
line (dashed); at this point, the cloud phase is at a triple point and
the properties of the shadow phase change discontinuously, here to a
dense phase B' richer in particles of species 1. Right: Schematic
cloud and shadow curves for this situation, showing the jump in the
shadow curve at the triple point. The cloud curve must be continuous
but generally has a kink at the triple point, where it switches
between different branches corresponding to
separate two-phase regions in the phase diagram.
\label{fig:three_phase}
}
\end{figure}
described above, let us return to the bidisperse case, but now with a
different phase diagram, shown in Fig.~\ref{fig:three_phase}.  This
phase diagram topology could occur in, for example, a binary liquid
whose constituent particles ``dislike'' each other; in addition to the
usual gas-liquid phase coexistence one can then also have demixing
into two liquids containing predominantly one of the particle species,
and three-phase coexistence between a gas and two such demixed
liquids. A dilute cloud phase may then begin to phase separate by
splitting off either one of these demixed liquids, depending on the
relative positions of the dilution line and the three-phase triangle
(see Fig.~\ref{fig:three_phase}). As temperature is varied, a corner
of the three-phase triangle may move through the dilution line. At the
temperature where this happens, the cloud phase is at a triple point,
and separates off infinitesimal amounts of {\em two different} shadow
phases; as $T$ is increased or decreased through the triple point, the
properties of the shadow phase therefore ``switch''
discontinuously. This implies that the
shadow curve will exhibit a jump discontinuity,
while the cloud curve remains connected as it must but will have a
kink where the shadow curve jumps (see
Fig.~\ref{fig:three_phase}). The coexistence curves also become more
complex: A parent may initially separate into two phases, but then
demix into three (or more, in the fully polydisperse case) as $T$
is lowered further; at the points where new phases appear, the
coexistence curves acquire new branches. For even lower
temperatures, the number of phases may increase yet further, or
decrease again. It is in fact an entirely open problem to predict from
the form of the free energy the maximum number of phases into which a
given polydisperse parent phases will separate.

One final new feature in the phase behaviour of polydisperse systems
is the possibility of encountering critical points of arbitrary order.
Such critical points are specified by a density distribution $\rhosig$
and a temperature $T$; their defining property is that, at those
parameters, a single phase separates into $n$ infinitesimally
different phases (on lowering $T$, for example).
Thus $n=2$ is an ordinary critical point, $n=3$ a
tricritical point and so on~\cite{Brannock91,SolWarCat01}. Since there
is no limit on the number of coexisting phases, it is intuitively
clear that there is also no upper limit on the order of critical points
that can occur in polydisperse systems. We will see a concrete example
of this below, for a simple model of a random copolymer blend.

\subsection{Formulating the general phase equilibrium problem}

The statistical mechanics of polydisperse systems is also known as
``continuous thermodynamics'' (see \eg~\cite{RatWoh91}). It is often useful
to separate off the ideal part of the free energy (density) explicitly
by writing (with $k_{\rm B}=1$)
\be
f([\rhosig]) = T\intsig \rhosig [\ln\rhosig -1] + \fexc([\rhosig])
\label{f_decomp}
\ee
This defines the excess free energy $\fexc$; both $f$ and $\fexc$ are
functionals of the density distribution $\rhosig$ (and also functions
of the externally fixed temperature, which I will not write
explicitly). The ideal part is the free energy of an ideal
polydisperse gas; it can be derived as the limiting form (up to an
irrelevant---infinite---constant term~\cite{SalSte82}) of the free
energy of an
ideal mixture of $M$ species, which is $T\sum_i \rho_i [\ln\rho_i -
1]$. Here the $\rho_i$ are again the number densities inside $M$ ``bins''
into which the range of $\sig$ has been partitioned, and the number of
bins is taken to infinity {\em after} the thermodynamic limit has been
performed. In an alternative derivation of the polydisperse limit, one
can assume from the start that all particles are genuinely different,
with $\sig$ sampled randomly from the normalized density distribution,
so that the number of distinct ``species'' is always $N$ and is taken
to infinity together with the system size. The two procedures give
equivalent results~\cite{Salacuse84}; an elegant derivation of the
ideal part of the free energy within the second approach was given by
Warren~\cite{Warren98}. Note that the first limit is physically more
plausible for many homopolymer systems (where there may only be
thousands or millions of species, with many particles of each) whereas
the second limit is more natural for colloidal materials (and also
some random copolymers) in which no two particles present are exactly
alike, even in a sample of macroscopic size.

From the free energy\eq{f_decomp}, the chemical potentials follow by
(functional) differentiation as
\be
\musig = \frac{\delta f}{\delta\rhosig} = T \ln\rhosig + 
\muexc(\sig), \qquad \muexc(\sig) = \frac{\delta \fexc}{\delta\rhosig}
\label{musig}
\ee
The pressure is, by analogy with\eq{Pi_monodisp},
\be
\label{Pi}
\Pi = -f + \intsig\musig\rhosig = T\rho_0 - \fexc +
\intsig\muexc(\sig)\rhosig
\ee
where
\be
\rho_0 = \intsig \rhosig
\label{rho_zero}
\ee
is the total number density of particles. Though not written
explicitly, both $\musig$ and $\Pi$ are functionals of $\rhosig$, and
ordinary functions of $T$. Note also that, in order avoid ``dimensional
crimes'' in the logarithms in $f$ and $\musig$,
eqs.\eqq{f_decomp}{musig}, one should really
divide the argument $\rhosig$ by a quantity with the same dimensions,
replacing \eg\ $\rhosig\to v_0\sig_0\rhosig$ (where $v_0$ and $\sig_0$
are chosen unit values of volume and of $\sig$) or
$\rhosig\to\rhosig/R(\sig)$ where $R(\sig)$ is a fixed density
distribution. As can be seen from\eq{musig}, however, any such
replacement would only add a $[\rhosig]$-independent term to the
chemical potentials
and so would not affect the predicted phase equilibria; we can therefore
proceed without it. (In the moment free energy method to be described
below, however, the fact that an arbitrary $R(\sig)$ can be chosen to
non-dimensionalize $\rhosig$ will be of crucial importance.)

Assume now that 
a given parent phase with density distribution $\rhonsig$ separates
into $P$ phases, labelled as before by $\al=1\ldots P$ and with
density distributions $\rhoalsig$. Then the chemical potentials
$\musig$ and the pressure $\Pi$ need to be equal in all phases, and
the total number of particles of each species $\sigma$ must be
conserved, implying that
\be
\sum_\al \ph\pa \rho\pa(\sig) = \rhonsig
\label{particle_cons}
\ee
where, as in\eq{discrete_cons}, $\ph\pa$ is the fraction of the
system volume occupied by phase $\al$.

From\eq{musig}, it follows that $\rhoalsig =
\exp[\beta\musig]\exp[-\beta\muexc\pa(\sig)]$, where $\beta=1/T$ and
$\muexc\pa(\sig)$ is the excess chemical potential of species
$\sigma$ in phase $\al$. Inserting into the particle conservation
law\eq{particle_cons}, one can eliminate $\exp[\beta\musig]$ and write
the density distributions in the coexisting phases as
\be
\rhoalsig = \rhonsig\,\frac{\exp[-\beta\muexc\pa(\sig)]}
{\sum_\gamma \ph^{(\gamma)} \exp[-\beta\muexc^{(\gamma)}(\sig)]}
\label{rhoalsig}
\ee
The $P$ unknown fractional phase volumes can then in principle be
determined from the equality of the pressure in all phases and from the
identity $\sum_\al \ph\pa = 1$. However, in\eq{rhoalsig} we have
achieved no more than a formal solution of the problem, since the
excess chemical potentials $\muexc\pa(\sig)$ are still functionals of
the unknown density distributions $\rhoalsig$ (so that, if these
functionals can be written as integrals, eq.\eq{rhoalsig} corresponds
in effect to $P$ coupled nonlinear integral
equations~\cite{GuaKinMor82}). 
%
%
Even if a valid solution for a phase split into $P$ phases could be
determined numerically, one would still need to verify that it is
thermodynamically stable, \ie\ that it gives the lowest possible total
free energy; this problem is exacerbated in a polydisperse system by
the potentially unlimited number of coexisting phases. In principle,
the criterion for stability is that no part of the free energy surface
``pokes through'' below the calculated tangent plane; equivalently, an
appropriately defined tangent plane distance~\cite{Michelsen82} needs
to be everywhere non-negative. Like $f$, however, the tangent plane
distance is a {\em functional} of $\rhosig$, so that a numerical
search over all its values is clearly impossible.

If one restricts oneself to finding not full phase splits, but just
the spinodal at which a given parent phase first becomes locally
unstable, one still faces a nontrivial task. A local instability
corresponds to a ``direction'' $\delta\rhosig$ in density
distribution space along which the curvature of the free energy
``surface'' vanishes (see \eg~\cite{SolWarCat01,Cuesta99}), such that
\be
\intsig \frac{\delta^2 f}{\delta\rh(\sig) \delta \rh(\sig')}
\delta\rh(\sig') = 0
\label{spinodal}
\ee
where the derivative is evaluated at the parent
$\rhonsig$. (I will not study here the subtle question of how, beyond
the approximate mean-field type models discussed below, free energies
can actually be defined in spinodal and unstable regions;
see \eg~\cite{FisZin98}.) The spinodal temperature $T$ can thus in
principle be found as the temperature where, coming from a region of
stability, this equation first has a nonzero solution $\delta\rhosig$.

\subsection{Truncatable free energies}

As explained above, predicting phase equilibria for a polydisperse
system with a completely generic free energy functional is next to
impossible. However, an important insight---later rediscovered by a
number of authors, and summarized in the most general terms probably
by Hendriks~\cite{Hendriks87,Hendriks88,HenVan92}---came from the
seminal work of Gualtieri {\em et al}~\cite{GuaKinMor82}: Significant
progress can be made for (model) systems with so-called
``truncatable'' free energies~\cite{SolWarCat01}. These are
characterized by an {\em excess} contribution $\fexc=
\fexc(\{\rho_i\})$ that depends only on a finite number, $K$ (say) of
generalized {\em \moms}
\be
\rho_i
= \intsig w_i(\sig) \rhosig
\label{rho_i}
\ee
of the density distribution $\rho(\sigma)$; for power-law weight
functions $w_i(\sigma)=\sigma^i$, the $\rho_i$ are conventional
moments. The term ``truncatable'' emphasizes that the number of \moms\
appearing in the excess free energy of truncatable models is finite,
while for a non-truncatable model the excess free energy depends on
all details of $\rhosig$, corresponding to an {\em infinite} number of
\moms. The class of polydisperse systems whose (at least
approximate) free energies are truncatable is surprisingly large; a
number of examples are given in Sec.~\ref{sec:systems} below. I will
normally assume that the total particle density $\rh_0$, corresponding
to the weight function $w_0(\sig)=1$, is included in the set of \moms.

For a truncatable system, the excess chemical potentials can be
written as
\be
\muexc(\sig) = \frac{\delta \fexc}
{\delta\rhosig} = \sum_i w_i(\sig) \muexc_i
\label{musig_trunc}
\ee
where the
\be
\muexc_i = \frac{\partial \fexc}{\partial \rho_i}
\ee
are excess moment chemical potentials. The density
distributions\eq{rhoalsig} in the different phases can thus be written
as
\be
\rhoalsig = \rhonsig \, \frac{\exp\left[\sum_i \lami\pa\wi\right]}
{\sum_\gamma \ph^{(\gamma)} \exp\left[\sum_i \lami^{(\gamma)}\wi\right]}
\label{rhoalsig_trunc}
\ee
where the $\lami\pa$ must obey
\be
\lami\pa = -\beta\muexc_i\pa + c_i
\label{lambda_i_exact}
\ee
The constants $c_i$ (common to all $P$ phases, with one for each \mom)
occur here since a common shift of all the $\lami\pa$ for any fixed
$i$ leaves the density distributions\eq{rhoalsig_trunc} unchanged. One
can fix this indeterminacy by, for example, setting all $c_i=0$, or
fixing all the $\lami\pa$ in one of the phases to be zero. Either way, we
have with\eq{lambda_i_exact} a set of $P\times K$ nonlinear
equations for the $P\times K$ parameters $\lami\pa$. At fixed values
of the $\ph\pa$, these equations are closed: From the
$\lami\pa$ one can find, via\eq{rhoalsig_trunc} and\eq{rho_i}, the
$\rho_i\pa$ and hence the $\muexc_i\pa$ (which, for a truncatable model,
are functions of the moment densities in the respective phase
only). The remaining $P$ parameters $\ph\pa$ are found again from
$\sum_\al \ph\pa=1$ and from the equality of the pressure
(using\eq{Pi})
\be
\Pi = T\rho_0 - \fexc + \sum_i \muexc_i \rho_i
\label{Pi_trunc}
\ee
in all phases. So the calculation of a phase split of a given parent
into $P$ phases requires, for a truncatable model, the solution of
$P(K+1)$ nonlinear coupled equations for the same number of variables.
Starting from a suitable initial guess, such a solution can, in
principle, be found by a standard algorithm such as
Newton-Raphson~\cite{PreTeuVetFla92}. (Generating an initial point
from which such an algorithm will converge, however, is a nontrivial
problem, especially when more than two phases coexist and/or many
\moms\ are involved.) There is also, for truncatable
systems, a well-defined way of checking whether a calculated phase
split is thermodynamically stable: Rather than over the
infinite-dimensional space of density distributions $\rhosig$, the
tangent plane distance now needs to be searched only over a
$K$-dimensional space, which is possible numerically using Monte Carlo
methods~\cite{SolWarCat01}.

If one is interested only in finding the cloud point for a given
parent distribution (rather than phase splits inside the coexistence
region), the problem becomes rather simpler. At the cloud point there
is coexistence between the parent $\rhonsig$, which still occupies all
of the system volume ($\ph\pn=1$), and $P$ shadow phases $\rhoalsig$
which are present in vanishingly small amounts ($\ph\pa=0$ for
$\alpha=1\ldots P$). In the generic situation there is only a single
shadow ($P=1$) but higher values of $P$ can occur, \eg\ $P=2$ at a
triple point (where, as discussed above, a cloud phase coexists with
two shadows). Using our freedom to choose the $\lami$ in one phase
to fix $\lami\pn=0$, we then have from\eq{rhoalsig_trunc} that the
shadow phase density distributions are given by
\be
\rhoalsig = \rhonsig \exp\left[\sum_i \lami\pa\wi\right]
\label{rhoalsig_cloud}
\ee
and that their $\lami\pa$ must obey
\be
\lami\pa = -\beta\muexc_i\pa + \beta\muexc_i\pn
\label{lami_cloud}
\ee
For an ordinary cloud point ($P=1$) there is then only one additional
equation, the equality of pressure between cloud (parent) and shadow,
and this fixes the cloud point temperature. (For larger $P$ the
pressure equalities give $P-1$ additional conditions on the parent
distribution $\rhonsig$; for $P=2$, for example, this condition
determines at what parent density the triple point occurs.)

Even more drastic simplifications occur, finally, in the spinodal and
critical point criteria for truncatable systems. In particular, it has
been shown that the spinodal criterion involves only the \moms\
$\rho_i$ of the parent phase, as well as its ``second order \moms''
$\rho_{ij}=\intsig\wi w_j(\sig) \rhonsig
$~\cite{SolWarCat01,Cuesta99,Hendriks88,HenVan92,GorIrvKen77,IrvGor81,BeeBerKehRat86,Warren99}.
This simplification is particularly useful if the \moms\ are ordinary
moments, \ie\ if the weight functions are simple powers $\wi=\sig^i$
($i=0\ldots K-1$); then the spinodal criterion only involves the
parent moments up to $\order(\sig^{2K-2})$. The condition for critical
points depends additionally on the third order \moms\ $\rho_{ijk}$
defined in the obvious way~\cite{SolWarCat01,IrvGor81,BeeBerKehRat87},
and generally one can show that the criterion for an $n$-critical
point will involve up to ($2n-1$)-th order \moms.

\section{Methods}
\label{sec:methods}

\subsection{Direct numerical solution}

For simple truncatable models involving only $K=1$ or 2 \moms, a
direct numerical solution of the phase equilibrium equations as given
above is often possible,
see \eg~\cite{RatWoh91,HenVan92,Bauer85,NesOlvCri93,Solc93,SolKon95,ClaMcLeiJen95};
I will review the results of some of these calculations below, in the
context of the various models that have been studied
(Sec.~\ref{sec:systems}). Apart from general purpose tools for solving
nonlinear coupled equations (see \eg~\cite{PreTeuVetFla92}), a number
of more specialized numerical techniques have been developed for this
purpose. Popular in particular in the chemical engineering literature
is the method of ``successive substitution''. This is based on an
iteration loop where at each iteration one first holds the excess
chemical potentials $\muexc\pa(\sig)$ (or, in a truncatable system,
the $\lami\pa$) fixed and finds the fractional phase volumes from the
conditions of pressure equality; this then determines the density
distributions, from which one can re-calculate the excess chemical
potentials and return to the beginning of the
loop~\cite{HenRos69,Prausnitz+80}. (Under conditions of constant
pressure rather than constant volume as considered here, the first
part of the iteration can be formulated as a minimization problem over
the $\ph\pa$~\cite{Michelsen94}.) Various accelerations and variants
of this method have been
proposed~\cite{Michelsen82,Michelsen82b,Michelsen86,Michelsen93}; a
serious disadvantage is, however, that the iteration can become
unstable and fail to converge~\cite{HeiMic95}. For the task of tracing
out cloud and shadow curves (rather than following the phase behaviour
of a given parent phase as external control parameters such as
temperature are varied) specialized techniques have also been
developed, see \eg~\cite{Michelsen94b}, with refinements for the
numerically often difficult regions around critical
points~\cite{NakDobInaYam99}.

\subsection{Binning and pseudo-components; method of moments}

For an approximate solution of the polydisperse phase equilibrium
problem, the most straighforward method is to ``bin'' the full density
distribution $\rho(\sigma)$ into a number of discrete
``pseudo-components'', whose densities are given by the density of particles
within the respective $\sig$-ranges. This then formally reduces the problem to
that of a finite mixture. The pseudo-components can be spaced evenly
across the $\sigma$-range, or chosen according to other ad-hoc
prescriptions. For simple functional forms of the parent distribution
$\rhonsig$ it has been suggested, for example, to locate the
pseudo-components at those $\sig$-values which would be used in a
Gaussian quadrature with $\rhonsig$ as the weight
function~\cite{CotPra85,ShiSanBeh87}. Some slight improvement in
accuracy is also possible by keeping track (to linear order) of
variations in the parent distribution across each bin~\cite{DroSch86}.
Whatever particular implementation is chosen, however, it is clear
that binning introduces uncontrolled systematic errors and also
becomes numerically unwieldy for large numbers of pseudocomponents.

A somewhat more systematic approach to allocating pseudo-components is
the method of ``$r$-equivalent
distributions''~\cite{GorIrvKen77,IrvKen82}. Here the parent
distribution $\rhonsig$ is replaced by a mixture of a finite number of
species whose $\sig$-values and densities are chosen to match
exactly the first $r$ moments of $\rhonsig$. If one is studying a
truncatable model (with power law weight functions), then since the
conditions for spinodals and critical points depend on only a finite
number of moments of $\rhonsig$, these points will be found exactly if
$r$ is chosen large enough. The results for actual phase splits,
however, including the onset of phase coexistence (cloud points and
shadows), will be only approximate.

An alternative (but still uncontrolled) approximation is the ``method
of moments''. This retains the
continuous range of $\sigma$ but fixes a parametric form for the
density distributions in all phases (\eg\ Gaussian or Schulz; the
Schulz distribution has the form $\rhosig \sim \sig^\alpha
e^{-\sig/\sig_0}$).
The free parameters specifying these distributions are then
found by solving the phase equilibrium equations approximately,
requiring particle conservation only for certain moments of the parent
density distribution $\rhonsig$ rather than all its
details~\cite{CotPra85,CotBenPra85,DuMan87,ManDuAnt89}. A similar
idea was used in Ref.~\cite{KinShoFes89} to reduce the problem of
finding cloud points and shadows to a set of (approximate) nonlinear
equations in a finite number of variables.

\subsection{Perturbative methods for nearly monodisperse systems}
\label{subsec:perturb}

For systems which are nearly monodisperse, one can pursue systematic
perturbation expansions around a monodisperse reference system. These
can never hope to capture qualitative polydispersity-induced changes in the
phase diagram, such as the appearance of new phases. However, they can
still give some important insights into the effects of ``weak''
polydispersity, predicting for example the trends in the fractionation
across coexisting phases, or whether polydispersity tends to narrow or
widen coexistence regions in the phase diagram.

The first of such perturbation theories was probably that of Gualtieri
{\em et al}~\cite{GuaKinMor82}. They assumed that the parent density
distributions consisted of a dominant monodisperse part ($\rhonsig
\sim \delta(\sig-\sig_0)$), with a small amount of polydisperse
material added. The overall fraction of polydisperse material was used
as the expansion parameter and therefore constrained to be small, but
there was no restriction on the width ($\sigma$-range) of the
polydisperse component. A number of other authors took a complementary
approach, assuming that the overall range of $\sig$-values in the
parent is narrow and expanding perturbatively in this small
width, often assuming a simple functional form for the parent
distribution such as a
Gaussian~\cite{KinShoFes89,KinMorLin83,BriGla84}.

More recently, Evans~\cite{EvaFaiPoo98,Evans99,Evans01,XuBau00} has
re-examined the perturbative approach and shown, in particular, that
the actual shape of the parent distribution is irrelevant (it can even
consist of a number of closely spaced $\delta$-peaks, corresponding to a
discrete mixture of very similar species) as long as it is
sufficiently narrow. In Evans' approach, it is useful to factor the
overall density out of all density distributions, decomposing $\rhosig
= \rn\probsig$ where $\probsig$ is the normalized
$\sig$-distribution ($\intsig \probsig = 1$). If the normalized parent
distribution $\probnsig$ is sufficiently narrow, with mean $\sav$,
then $\eps=(\sig-\sav)/\sav$ will be small in all coexisting phases;
it is then convenient to switch from $\sig$ to $\eps$ as the
polydisperse attribute. Evans now assumes that the excess free energy
(density) of an arbitrary phase with density distribution
$\rho(\eps)=\rn\prob(\eps)$ can be expanded systematically as
\be
\beta\fexc = \fexc_{\rm m}(\rn) + \lav\eps\rav A(\rn) + \langle
\eps^2\rangle B(\rn) +
\lav\eps\rav^2 C(\rn) + \order(\eps^3)
\ee
where $\lav\eps\rav=\int\!d\eps\,\eps\,\prob(\eps)$ and similarly for
$\lav\eps^2\rav$, and $\fexc_{\rm m}(\rn)$ is the excess free energy
of a monodisperse reference system (with $\eps=0$, \ie\ $\sig=\sav$
for all particles). The coefficients $A$, $B$ and $C$ are unspecified
functions of the overall density $\rn$. From this very generic form a
number of elegant results follow. For example, for the normalized
$\eps$-distribution in a phase $\al$ coexisting with one or more other
phases, one finds
\be
\prob\pa(\eps) =
\prob\pn(\eps)\left[1-\eps\left(\frac{A\pa}{\rn\pa}-\frac{1}{\rn\pn}
\sum_\beta \ph\pb A\pb\right) \right] + \order(\eps^2)
\label{evans_peps}
\ee
where the coefficients $A\pa\equiv A(\rn\pa)$ can be evaluated at the densities
of the coexisting phases in either the monodisperse reference system
or the actual 
polydisperse system, the difference contributing only to the neglected
$\order(\eps^2)$ terms. Taking the first moment of\eq{evans_peps}, one has
for the difference of $\lav\eps\rav$ in two coexisting phases
\be
\lav\eps\rav\pa-\lav\eps\rav\pb = - s^2
\left(\frac{A\pa}{\rn\pa}-\frac{A\pb}{\rn\pb}\right)
\label{evans_universal}
\ee
where $s$, defined through
\be
s^2 = \int\! d\eps\, \eps^2\, \prob\pn(\eps)  = \intsig 
\left(\frac{\sig-\sav}{\sav}\right)^2 \probnsig
\label{s_def}
\ee
is the standard deviation of the parent distribution normalized by its
mean, often simply called the polydispersity. The ``universal
fractionation law''~\cite{EvaFaiPoo98,Evans99} of eq.\eq{evans_universal}
states that the difference in the mean of the particle sizes (or
whatever polydisperse attribute $\eps$ measures) in coexisting phases
is directly proportional to the {\em variance} of the parent
distribution. The result is valid for arbitrary (narrow) parent
distributions, including non-smooth ones.  Results for the
polydispersity-induced shifts of phase boundaries relative to the
monodisperse reference system can also be derived, and are again found
to be proportional to the variance $s^2$ of the parent distribution
(rather than, as one might have naively expected, to its standard
deviation $s$). In the region near critical points, the perturbative
expansion for the phase boundaries breaks down, since polydispersity
generally shifts the location of the critical point to a different
temperature; at the critical point of the monodisperse reference
system, a polydisperse system will thus show either
non-critical phase coexistence (between non-identical phases), or no
phase separation at all. Nevertheless, the approach is useful,
particularly if one is interested in questions such as whether
polydispersity will lead to a widening or a narrowing of the
coexistence gap in any given system. It also generalizes
straightforwardly to the case of several polydisperse attributes,
where $\eps$ becomes a vector-valued variable~\cite{Evans01}.

\subsection{Moment free energy method}

As pointed out above, even for truncatable models the numerical
solution of the phase equilibrium conditions can be an extremely
difficult numerical problem. Furthermore, the nonlinear phase
equilibrium equations permit no simple geometrical interpretation or
qualitative insight akin to the familiar rules for constructing phase
diagrams from the free energy surface of a finite mixture. To address
these two disadvantages, one can construct a so-called ``moment free
energy''~\cite{SolWarCat01,Warren98,SolCat98}. This takes the above
insights for truncatable systems further, by showing that 
a simplification similar to that for the phase equilibrium conditions
exists also on the level of the free energy itself.

There are (at least) two approches to constructing the moment free
energy; I describe here the so-called projection
method~\cite{SolCat98}. The starting point is the
decomposition\eq{f_decomp} for the free energy of truncatable systems
\be
f = T \intsig \rhosig \left[\ln \frac{\rhosig}{R(\sigma)}
-1\right] + \fexc(\{\rho_i\})
\label{free_en_decomp}
\ee
In the first (ideal) term of\eq{free_en_decomp}, a dimensional
factor $R(\sigma)$ has been included inside the logarithm; while this has
no effect on the exact thermodynamics (see above), it will play a central
role below. 

To motivate the construction of the moment free energy, one
can argue that the most important \moms\ to treat correctly
in the calculation of phase equilibria 
are those that actually appear in the excess free energy
$\fexc(\{\rho_i\})$.  Accordingly one divides the infinite-dimensional
space of density distributions into two complementary subspaces: a
``moment subspace'', which contains all the degrees of freedom of
$\rhosig$ that contribute to the \moms\ $\rho_i$, and a ``transverse
subspace'' which contains all remaining degrees of freedom (those that
can be varied without affecting the chosen \moms\ $\rho_i$). Physically,
it is reasonable to expect that these ``leftover'' degrees of freedom
play a subsidiary role in the phase equilibria of the system, a view
that can be justified {\em a posteriori}. Accordingly, one now allows
violations of the lever rule, so long as these occur {\em solely in
the transverse space}. This means that the phase splits calculated
using this approach obey particle conservation for the \moms, but are
allowed to violate it in other details of the density distribution
$\rhosig$. These ``transverse'' degrees of freedom are instead chosen
so as to minimize the free energy: they are treated as ``annealed''.
Because the excess free energy depends (for a truncatable system) only
on the set of \moms, one therefore has to minimize the ideal part of
the free energy over all distributions $\rhosig$ with fixed \moms\
$\rho_i$. This yields
\be
\rhosig=R(\sig)\exp\left[\sum_i \lami\wi\right]
\label{family}
\ee
where the Lagrange multipliers $\lami$ are chosen to give the desired
\moms
\be
\rho_i = \intsig\wi\,R(\sig)\exp\left[\sum_j \lam_j\w_j(\sig)\right]
\label{moments_from_lambda}
\ee
The corresponding minimum value of $f$ then defines the {\em moment free
energy} as a function of the \moms\ $\rho_i$:
\be
\fmom(\{\rho_i\}) = T \left(\sum_i \lami\rho_i - \rn \right)  +
\fexc(\{\rho_i\})
\label{fmom}
\ee
Since the Lagrange multipliers are (at least implicitly) functions of
the \moms, the moment free energy depends only on the set of
\moms. These can now be viewed as densities of ``quasi-species'' of
particles, allowing for example the calculation of ``moment chemical
potentials''~\cite{SolWarCat01}
\be
\mu_i= \frac{\partial\fmom}{\partial\ri} =
T\lami + \frac{\partial\fexc}{\partial\ri} =  T\lami + \muexci
\label{mom_chem_pot}
\ee
and the corresponding pressure
\be
\Pi = \sum_i \mu_i \ri - \fmom = T\rn + \sum_i \muexci \ri - \fexc
\label{mom_osmotic_pressure}
\ee
(which for truncatable systems is identical to the exact
expression\eq{Pi_trunc}).
A finite-dimensional phase diagram can thus be
constructed from $\fmom$ according to the usual tangency plane rules,
ignoring the underlying polydisperse nature of the system.  Obviously,
though, the results now depend on $R(\sig)$ which is formally a
``prior distribution'' for the free energy minimization.  {\em
Geometrically}, its effect is to tilt the free energy surface before
it is ``projected'' onto the moment subspace; this point of view is
explained in detail in Ref.~\cite{SolWarCat01}. To understand the
influence of $R(\sig)$ {\em physically}, one notes that the moment
free energy is simply the free energy of phases in which
the density distributions $\rhosig$ are of the form\eq{family}. The
prior $R(\sig)$ determines which distributions lie within this
``family'', and it is the properties of phases with these
distributions that the moment free energy represents. To ensure
that the parent phase is contained in the family, one chooses its
density distribution as the prior, $R(\sig)=\rhonsig$; the moment free
energy procedure will then be {\em exactly valid} whenever the density
distributions {\em actually arising} in the various coexisting phases
of the system under study {\em are members of the corresponding
family}
\be
\rhosig=\rhonsig\exp\left[\sum_i \lami\wi\right]
\label{pfamily_precap}
\ee
This condition holds whenever all but one of a set of coexisting
phases are of infinitesimal volume compared to the majority phase, as
can be seen explicitly from\eq{rhoalsig_cloud}. Accordingly, the moment free
energy yields {\em exact} cloud point and shadow curves. (And the
exact conditions\eq{lambda_i_exact} are seen, with the help
of\eq{mom_chem_pot}, to
express precisely the requirement of equal moment chemical potentials in
all phases.) Similarly, one can show that spinodals and critical
points of any order are found exactly~\cite{SolWarCat01}. For
coexistences involving finite amounts of different phases the moment
free energy only gives approximate results, since different density
distributions from the family\eq{pfamily_precap}, corresponding to two
(or more) phases arising from the same parent $\rhonsig$, do not in
general add to recover the parent distribution itself. Moreover,
according to Gibbs' phase rule, a moment free energy depending on $K$
\moms\ will not normally predict more than $K+1$ coexisting phases,
whereas we know that a polydisperse system can in principle separate
into an arbitrary number of phases. Both of these shortcomings can be
overcome by including extra \moms\ within the moment free energy; this
does not affect any of the exactness statements above but
systematically increases the accuracy of any calculated phase
splits~\cite{SolWarCat01}. This idea can be further refined by
choosing the weight functions of the extra moments adaptively, which
allows the properties of the coexisting phases to be predicted with in
principle arbitrary accuracy~\cite{ClaCueSeaSolSpe00}.

The moment free energy method (or moment method for short) thus
restores to the problem of polydisperse phase equilibria much of the
physical and geometrical insight available from the thermodynamics of
finite mixtures. It also leads to computationally efficient
procedures; in particular, its numerical implementation can handle
coexistence of more than two phases with relative ease compared to
previous
approaches~\cite{Michelsen82,Hendriks88,Michelsen94,Michelsen86,%
Michelsen94b,CotPra85,ShiSanBeh87,IrvKen82,CotBenPra85,KinShoFes89,%
Michelsen87,Michelsen86b}.

\section{Applications to (model) systems}
\label{sec:systems}

\subsection{Polymers I: Flory-Huggins theory for homopolymers}

\begin{figure}
\begin{center}
\epsfig{file=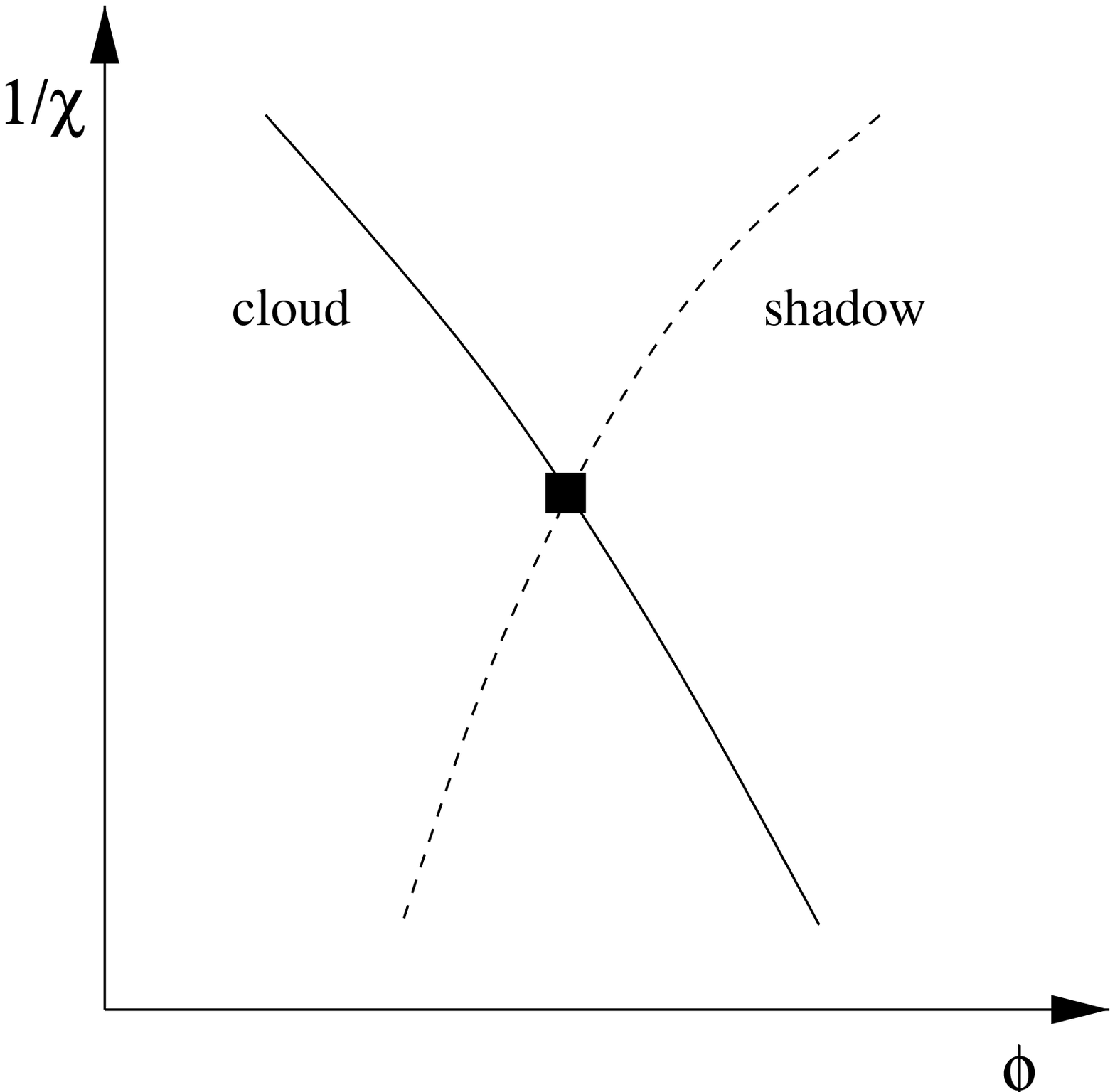,width=0.25\textwidth}\hspace*{3mm}%
\epsfig{file=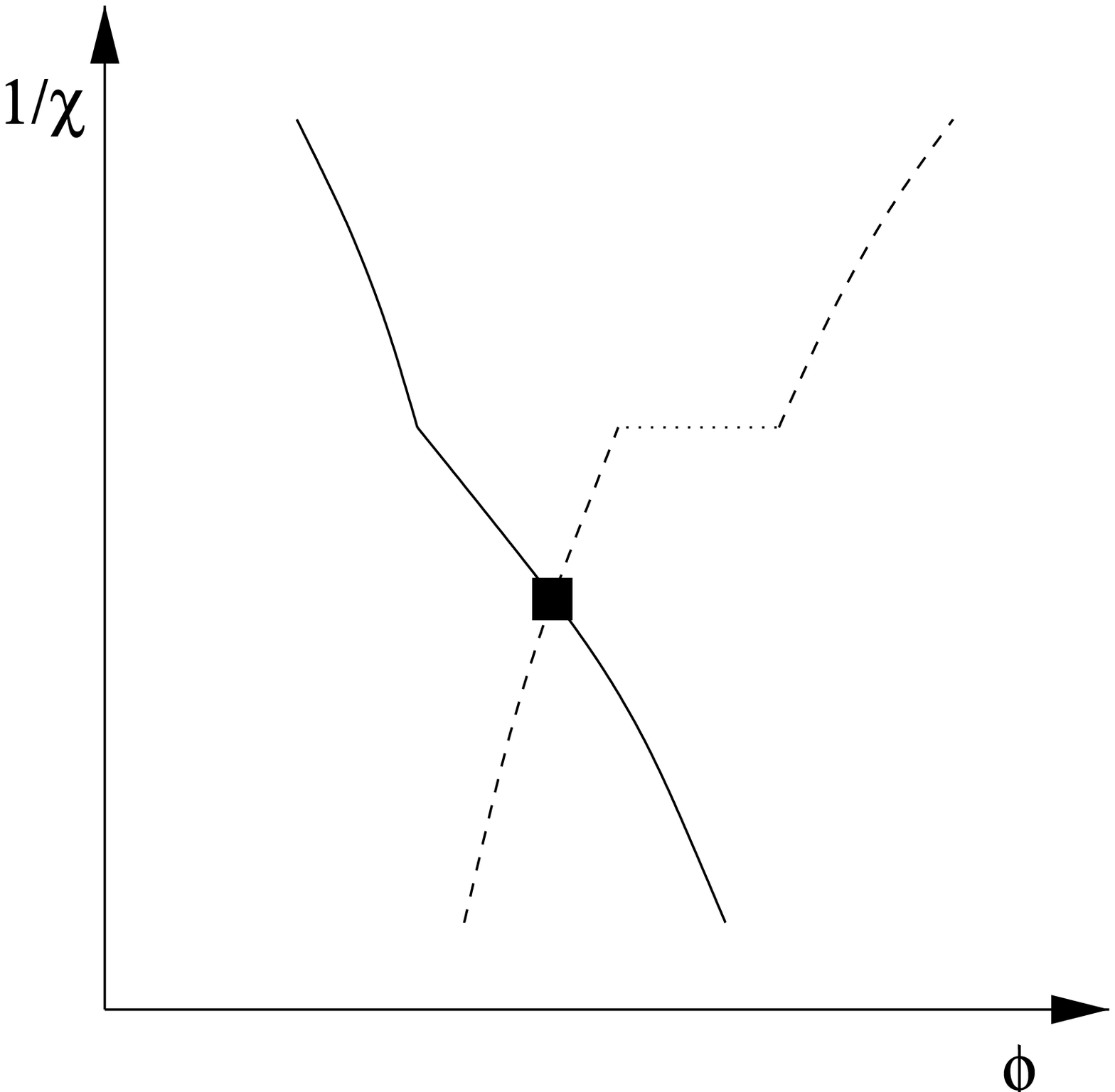,width=0.25\textwidth}\hspace*{3mm}%
\epsfig{file=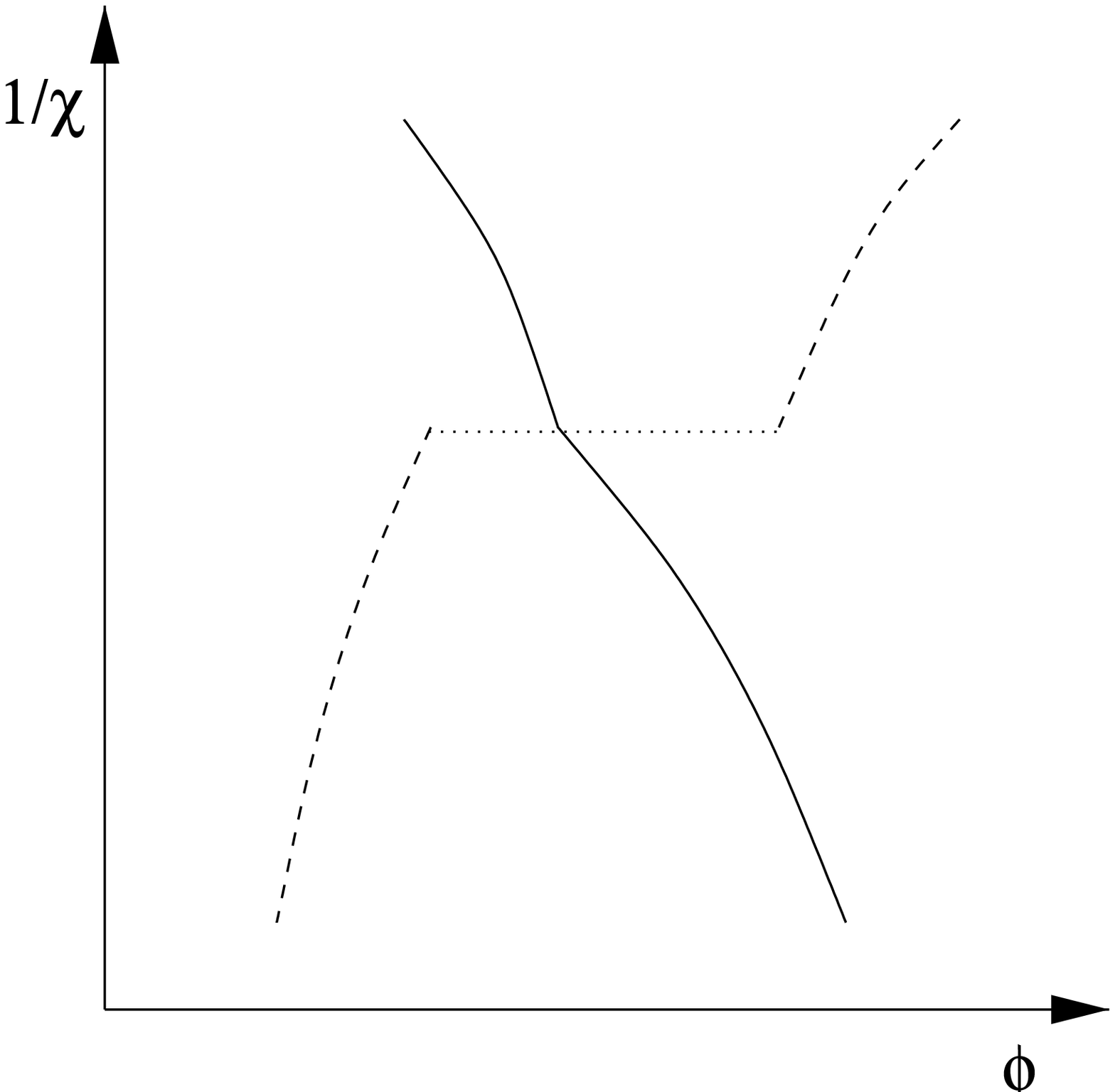,width=0.25\textwidth}
\end{center}
\caption{Qualitative cloud and shadow curves for
homopolymers with chain length polydispersity. Only the region around the
critical point (which is marked by a rectangle) is shown; cloud curves are
solid and shadow curves dashed. As is conventional, the polymer volume
fraction (rather than number density) is used on the $x$-axis to
specify the overall dilution of the system; on the $y$-axis, since
$\chi$ is measured in units of $k_{\rm B}T$, $1/\chi$
is essentially a dimensionless temperature. Left: The conventional
phase diagram topology, which is found for monodisperse or weakly
polydisperse systems. Middle: As the length polydispersity increases,
a triple point can occur where the shadow curve has a jump
discontinuity. Right: For even more pronounced polydispersity, the
triple point may prevent the crossing of cloud and shadow curves, thus
making the critical point inaccessible.
\label{fig:polydisp_FH}
}
\end{figure}

Flory-Huggins theory~\cite{Flory53} is a simple but remarkably
successful approximate theory describing the thermodynamics of polymer
solutions and blends. It was derived in the
1940's~\cite{Flory44,ScoMag45,Scott45} and extended very early on to
include polydispersity in the lengths of the polymer chains. For
clarity I will use $L$ here rather than $\sig$ for the polydisperse
attribute, so that $\rhoL\ dL$ will be the number density of polymers with
lengths in the range $[L,L+dL]$. The excess free energy of
polydisperse Flory-Huggins theory for homopolymers (which
contain only one type of monomer) is then
\be
\fexc = (1-\ro)\ln(1-\ro) + \chi\ro(1-\ro)
\label{FH_def}
\ee
where $w_1(L)=L$ and I have set $k_{\rm B} T=1$. I have also chosen
the volume of a solvent molecule as the unit volume, and assumed for
simplicity that this is equal to the volume of a monomer (or polymer
``segment''); $\ro$ is then simply the volume fraction of polymer. The
first term in $\fexc$ is minus the entropy of the solvent and always
leads to an increase in free energy when phase separation occurs. The
second term, on the other hand, reflects the interactions of the
monomers with each other and with the solvent, with $\chi$ measuring the
effective monomer-monomer attraction in units of $k_{\rm B} T$. When
(as $T$ is lowered) $\chi$ becomes sufficiently large, this attraction
causes a phase separation into a polymer-rich and a polymer-poor
phase; in the monodisperse case, this is the only phase separation
that occurs. Tompa~\cite{Tompa49,Tompa56}, however, realized that
already bidisperse polymer solutions can exhibit three phase
coexistence as soon as the ratio of the two different chain lengths is
larger than around ten; this then produces a kink in the cloud curve
and a jump in the shadow curve as discussed in
Sec.~\ref{sec:basics}. Solc~\cite{Solc70} realized that, in fact, rather
intricate phase diagram topologies can occur: As sketched in
Fig.~\ref{fig:polydisp_FH}, the occurrence of the three-phase
coexistence can end up ``removing'' the critical point from the cloud
curve, by shifting it onto a metastable or unstable branch of the
cloud curve where it is no longer accessible.

Of course, three phase coexistence will be observed not only for the
parent density at which the cloud curve has its kink (\ie, directly at
the triple point), but also for a range of temperatures and densities
around this point. Not unexpectedly, the maximum temperature interval
over which three phase coexistence can be observed (for an
appropriately chosen composition of the system) becomes wider as the
lengths of the two polymer species becomes more disparate; conversely,
it can shrink to zero as they become comparable. Where this happens,
one gets a tricritical point, as was first realized by Solc {\em et
al.}~\cite{SolKleKon84} and later confirmed
experimentally~\cite{NakDobInaYam99,DobNak86,NakDob86,SunWid88,SheSmiKnoSco90}.
For mixtures of more than two polymer species with appropriately tuned
length distributions, higher order critical points can also
occur~\cite{SolBat85}. Finally, more complicated phase separation
sequences including even re-entrant features are possible; for a
specific solution of a mixture of three different polymer species, for
example, a sequence of one $\to$ three $\to$ two $\to$ three phases
was observed on lowering temperature~\cite{SuzDobMikYamNak00}.

All the results above were for mixtures of a small number of distinct
polymer species whose chain lengths were assumed to be sharply
defined. For truly
polydisperse systems, the first numerical calculations of phase
equilibria were probably those of Koningsveld and
Stavermann~\cite{KonSta67,KonSta67b,KonSta68,Koningsveld69}, with
an emphasis on using fractionation effects to generate phases with a
narrow distribution of chain lengths. Solc~\cite{Solc70,Solc75}
realized later that three-phase coexistence is quite
generic in distributions of chain lengths which have ``fat tails''
(which means, in this context, that they decay more slowly than
exponential with $L$ for large lengths).
%
%
In such systems he predicted the critical point to be always
``hidden'', corresponding to the most extreme polydisperse case
sketched in Fig.~\ref{fig:polydisp_FH}, and this was later
confirmed experimentally~\cite{DelPooDie83}. 

In fact, ``fat-tailed'' parent distributions give rise to quite subtle
behaviour, in particular for the cloud and shadow curves. To explain
how this arises, I will paraphrase Solc's
arguments~\cite{Solc70,Solc75} here. Flory-Huggins theory for
homopolymers gives, as can be seen from\eq{FH_def}, a truncatable
free energy with a single \mom\ $\ro$, with excess moment chemical potential
\be
\muexc_1 = -1 - \ln(1-\ro) + \chi(1-2\ro)
\label{muexc_FH}
\ee
and osmotic pressure
\be
\Pi = \rn + \ro \muexc_1 - \fexc = \rn - \ro - \ln(1-\ro) - \chi\ro^2
\label{Pi_FH}
\ee
The solvent entropy leads to a positive contribution $- \ro -
\ln(1-\ro)$ to $\Pi$, acting against increases in polymer volume
fraction $\ro$, while the monomer-monomer attraction gives the
negative term $-\chi\ro^2$ favouring large values (but $<1$) of $\ro$.

Consider now the shadow phase coexisting with the parent at the cloud
point. From\eq{lami_cloud}, its density distribution has the form
$\rh\po(L) = \rh\pn(L) \exp(\lamo\po L)$. If we drop the superscript
on the shadow phase properties and abbreviate $\lam\equiv\lamo\po$
this is written simply as
\be
\rhoL = \rh\pn(L)\exp(\lam L)
\label{FH_shadow}
\ee
and the condition on $\lam$---sometimes called the ``separation
parameter'' in the polymer literature---is, from\eq{lami_cloud}
\be
\lami = \beta\muexc_1\pn - \beta\muexc_1
\ee
The value of $\chi$ at the cloud point, finally, can be found from the
pressure equality $\Pi=\Pi\pn$.

Eq.\eq{FH_shadow} shows clearly why a slower-than-exponential decay of
the parent distribution for large $L$ will lead to unusual effects: A
positive value of $\lam$ causes all moments of the shadow phase
distribution $\rhosig$ to diverge. In fact, to get well-defined
results one needs to impose---as is physically reasonable---a cutoff
on the parent distribution at some large length $\lmax$, and then
consider the limit\footnote{%
The strict limit $\lmax\to\infty$ is of course unrealizable
physically, but useful as a mathematical device for highlighting the
effects of large $\lmax$.}%
\ of large $\lmax$. Cloud-shadow pairs with negative
$\lam$---corresponding to a dense cloud phase and a more dilute
shadow---will be only very weakly affected by the value of $\lmax$,
since no diverging integrals occur even for $\lmax\to\infty$.  For
positive $\lam$, on the other hand, one needs to consider carefully
the dependence of $\lam$ on $\lmax$. A first possibility is that
$\lam$ has a nonzero limit for $\lmax\to\infty$. But then the integral
for the shadow's polymer volume fraction $\ro = \intL L
\,\rh\pn(L)\exp(\lam L)$ will diverge unless the parent (cloud phase)
density $\rn\pn$ converges to zero such as to give a limiting $\ro<1$.
The polymer density $\rn$ of the shadow will then converge to zero,
since the shadow phase is dominated by the longest polymer chains and
thus $\rn \sim \ro/\lmax\to 0$. From\eq{Pi_FH}, the osmotic
pressure of the shadow phase is then
\be
\Pi = -\ro -\ln(1-\ro) -\chi\ro^2
\ee
The parent, on the other hand, has $\Pi\pn=0$ because of its vanishing
polymer density $\rn\pn$ (and hence polymer volume fraction $\ro\pn$). The
pressure equality thus gives $\Pi=0$, or
\be
\chi = \frac{-\ro -\ln(1-\ro)}{\ro^2}
\ee
Remarkably, this result for the shadow curve (expressed as $\chi$ vs
shadow polymer volume fraction $\ro$) is {\em universal}, \ie\
independent of any features of the parent $\rh\pn(L)$ except the
presence of a fat tail~\cite{Solc75}. The resulting phase coexistence
is rather peculiar: At vanishingly small polymer density (and volume
fraction), the parent splits off a shadow with a finite polymer
volume fraction, and made up of only the very longest chains in the
parent. Both phases have vanishingly small pressure; in the shadow,
this is achieved by an exact balance between the positive (repulsive)
and negative (attractive) contributions to $\Pi$.

\begin{figure}
\begin{center}
\epsfig{file=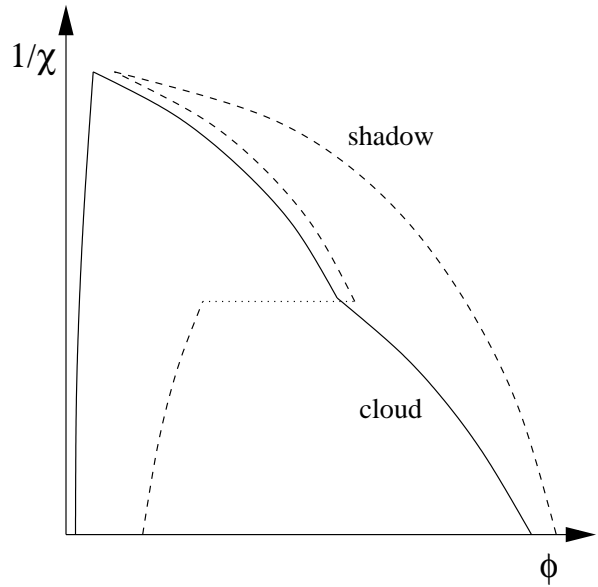,width=0.5\textwidth}
\end{center}
\caption{Sketch of cloud and shadow curves for a homopolymer with
a fat-tailed length distribution. In the limit where the cutoff $\lmax$
on chain lengths becomes very large, the low density part of the cloud
curve becomes vertical, while for polymer volume fractions above zero
but below the triple point the cloud and shadow curves approach each
other and eventually coincide.
\label{fig:fat_tail_FH}
}
\end{figure}

The reasoning so far gives the onset of phase coexistence which occur
as the parent density $\rn\pn$ is increased at fixed (sufficiently
large) $\chi$. If one wants to find instead the cloud point at fixed
nonzero $\rn\pn$, caused by an increase in $\chi$,
Solc~\cite{Solc70,Solc75} showed that one has to assume that $\lam\to
0$ as $\lmax\to\infty$. This means that the cloud and shadow are
``almost critical'': For small chain lengths $L$ (of order the parent
average), their density distributions are essentially identical; the
shadow only differs from the parent 
in that it contains a larger fraction of the longest
chains ($L\approx \lmax$). As a consequence, when represented in a
$\chi$ vs $\ro$ plot, the cloud and shadow curves actually {\em
coincide} in the limit $\lmax\to\infty$; their functional form is
again universal and given by $2\chi =
1/(1-\ro\pn)=1/(1-\ro)$. However, due to the contribution
from the longest chains, all moments $\rh_n = \intL L^n \rh(L)$ of the
shadow with $n>2$ actually {\em diverge} with $\lmax$; cloud and
shadow curves plotted as $\chi$ vs $\rh_2$, for example, would
therefore be extremely different (by an infinite amount in the limit).
A sketch summarizing the
overall shape of the cloud and shadow curves for polymers with
fat-tailed length distributions is shown in
Fig.~\ref{fig:fat_tail_FH}.

The above considerations are not as academic as they may seem; 
log-normal length distributions, for example, have fat tails as
defined above and occur 
frequently in polymer processing. For branched polymers, length
distributions with (even fatter) power law tails arise naturally, and
lead to similar phenomena~\cite{ClaMcLeiJen95}.

Flory-Huggins theory is by its nature a mean-field theory; as
described above, it is nevertheless rather successful at capturing the
effects of length polydispersity on polymer phase behaviour.  Close to
critical points, deviations will occur; even there, however,
polydispersity has been shown to have nontrivial effects. For example, while
monodisperse polymers display critical behaviour of the Ising
universality class, the critical exponents are modified nontrivially
by polydispersity, due to the presence of the large number of
conserved densities which act as ``hidden
variables''~\cite{KitDobYamNakKam97}.

\subsection{Polymers II: Random copolymers}
\label{sec:copol}

Flory-Huggins theory can also be applied to copolymers, which are made
up of random sequences of two types (A and B, say) of monomer.  Define
$\sig$ as the difference between the fractions of A- and B-monomer on a
chain, such that $\sig\in[-1,1]$. One can then have polydispersity in
the polymer chain lengths, $L$, as well as in the chemical
chain compositions, $\sig$, and so the system is described by a density
distribution $\rho(L,\sig)$. In the same units as for the homopolymer
case, Flory-Huggins theory then gives for the excess free energy
\be
\fexc = \frac{1}{\ls}(1-\rho_1)\ln(1-\rho_1) - \chi\rho_1^2
 - \chi' \rho_2^2 - \chi'' \rho_1\rho_2
\label{FH_copol}
\ee
Two \moms\ now appear, defined by the weight functions $w_1(L,\sigma)
= L$ and $w_2(L,\sigma) = L\sigma$. I have also included the
generalization to a polymeric solvent here, with chain length $\ls$.
As before, $\chi$ measures the effective monomer-monomer
attraction, but two additional parameters now appear: $\chi'$ favours
A-B demixing, and $\chi''$ accounts for any asymmetry in the
interactions between solvent and monomers A and B, respectively. In
the homopolymer case, where A and B are identical, one has
$\chi'=\chi''=0$ and then retrieves the expression\eq{FH_def} as
expected (up to 
the term $\chi\ro$, which is linear in density and so irrelevant for
the phase behaviour).

The case of polydisperse $L$ {\em and} $\sigma$ is rather complex, so it is
easiest to extract the copolymer-specific effects first by assuming
that only the chemical composition $\sigma$ is polydisperse while the
chain length $L$ is monodisperse. One can then replace
$\rho(L,\sigma)\to \rho(\sigma)$, and the \moms\ $\rn$ and $\ro$
become trivially related according to $\ro=L\rn$; similarly, $\rt$
becomes $L$ times the first moment (w.r.t.\ $\sig$) of $\rhosig$.

\begin{figure}[p]
\begin{center}
\epsfig{file=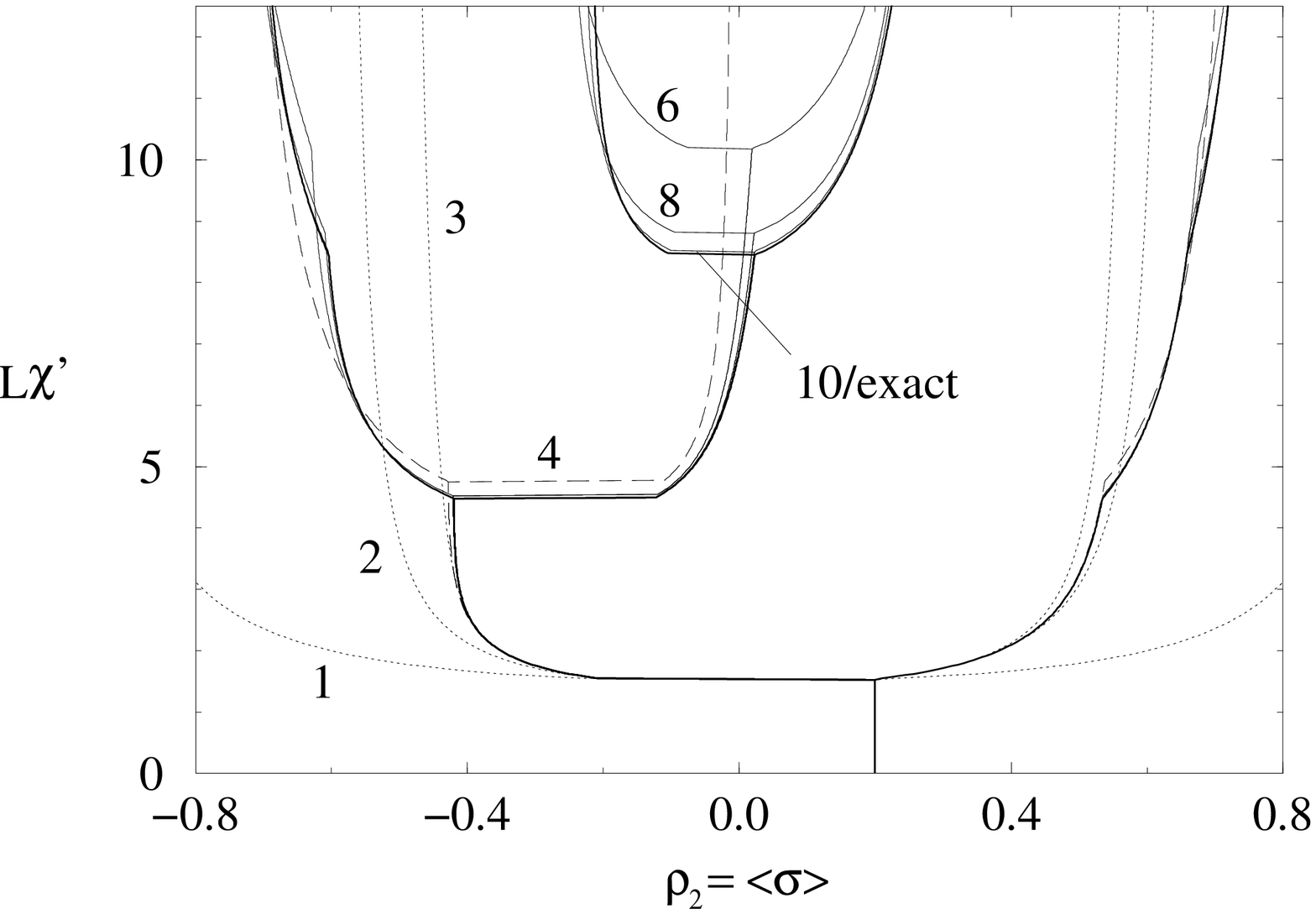,width=0.6\textwidth}

\hspace*{-2mm}\epsfig{file=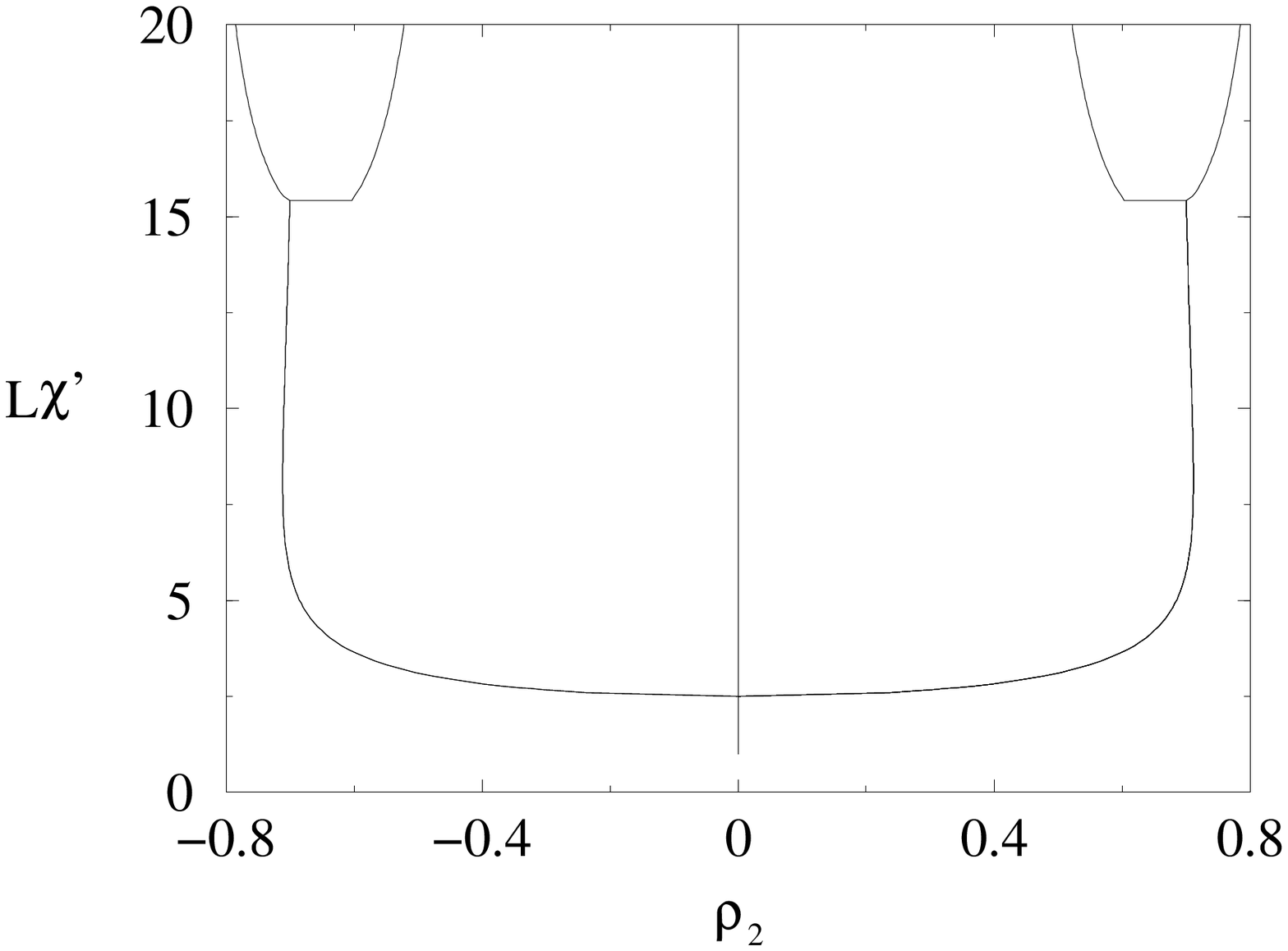,width=0.62\textwidth}
\end{center}
\caption{Top: Example of a demixing cascade in a random copolymer
blend (\ie\ without solvent), as calculated from Flory-Huggins
theory~\cite{SolWarCat01}. Shown are, for a single parent phase,
the average values of
$\rho_2=\lav\sig\rav$ in the coexisting phases as $\chi'$ is increased,
corresponding to temperature being decreased. Note how more and more
coexisting phases appear, producing new branches of the coexistence curve
(connected by horizontal lines to guide the eye).
The different curves are calculated using
the moment free energy method, and labelled by the number $K$ of
moments retained in the description. While the results for the cloud
point and shadow are exact even with the smallest $K$ ($K=1$), the
predictions in the coexistence region approach those of an exact
calculation as $K$ is increased, and are indistinguishable for
$K=10$. Bottom: With solvent added to the system, parents with
appropriately chosen polymer volume fraction exhibit tricritical
points, where separation into three infinitesimally different
phases occurs~\cite{SolWarCat01}. (Only the---essentially exact---results for
the largest value of $K$ are shown.)
\label{fig:dense_copol}
}
\end{figure}

To simplify even further, one can assume that there is no solvent in
the system, constraining the polymer volume fraction to
$\ro=L\rn=1$. The normalized $\sig$-distribution is then
$\probsig=L\rhosig$, $\rho_2$ reduces to the average of $\sig$, and
the excess free energy becomes $\fexc = -\chi'\rho_2^2 = -\chi'
[\intsig \sig\probsig]^2$ up to constants and irrelevant linear
terms. By adding back some linear terms (and exploiting the fact that
$\intsig \probsig = 1$), one can also write this excess free energy as
\be
\fexc = \half \chi'\int\!d\sigma\, d\sigma'\, (\sigma-\sigma')^2
\probsig\prob(\sig')
\label{FH_dense_copol}
\ee
which shows quite transparently the mechanism of phase separation in
this system: Phases that contain a spread of different $\sigma$ can
always lower their excess free energy by fractionation; as temperature
is lowered, this effect dominates the corresponding loss of entropy of
mixing in the ideal part of the free energy, and one expects
separation into an ever-increasing number of phases\footnote{%
This phase separation occurs at values of $\chi'$ of order $1/L$,
rather than of order unity; the reason for this is that the ideal part
of the free energy is $\intsig \rhosig[\ln\rhosig - 1] = L^{-1}\intsig
\probsig\ln\probsig$ ($+$ linear terms).
}
. This remarkable
behaviour has indeed been found in numerical
calculations~\cite{Bauer85,NesOlvCri93,Solc93}; an example is shown in
Fig.~\ref{fig:dense_copol}(top). Equally remarkably, one can show that
in even such a very simple model system, critical points of arbitrary
order can occur (though the required fine-tuning of the parent density
distribution would likely make the experimental observation of 
critical points of higher 
order than tricritical very difficult)~\cite{SolWarCat01}.

For the case where solvent is present in the system, but the chain
lengths $L$ are still monodisperse, phase coexistences have been
calculated in Ref.~\cite{SolWarCat01}. To simplify matters, the
solvent was assumed to be polymeric and to have the same chain length as
the copolymer ($\ls=L$). The interaction parameters $\chi$ and
$\chi''$ were also taken to be zero, with only $\chi'$ being nonzero;
under these assumptions, the solvent acts exactly like a random
copolymer chain with $\sig=0$, \ie\ composed of equal numbers of A and
B monomers. Even for this simple scenario,
all symmetric (under $\sig\to-\sig$)
parent density distributions $\rhonsig$
show a tricritical point at some value of the overall polymer
density (see Fig.~\ref{fig:dense_copol}(bottom)). This effect
generalizes that found in a simpler bidisperse case,
where the ``copolymer mixture'' only contains the pure A and
pure B homopolymers~\cite{Leibler81,BroFre90}. The fully 
polydisperse case is nevertheless richer since it also allows 
critical points of higher order than tricritical.

Finally, the most general case of joint polydispersity in lengths $L$
and chemical compositions $\sig$ has been treated by a number of
authors (see \eg~\cite{RatWoh91} for an extensive review). In early
work an incorrect form for the excess free energy was
used~\cite{RatKehBro85,RatKehBroSch86}; the correct form is the one
given above in eq.\eq{FH_copol}~\cite{Solc93,RatBroKeh89}. While the
calculation of phase splits inside the coexistence region remains an
open problem, the simpler 
cloud and shadow curves have been obtained for a number
of scenarios. One interesting feature is that even for Schulz
distributions of chain lengths, triple points and the associated kinks
in the cloud curve are predicted (and observed
experimentally~\cite{RatKehBroSch86}). These are clearly due to the chemical
polydispersity since homopolymers with Schulz distributions of chain length
never exhibit triple points~\cite{Solc70,Solc75}.

\subsection{Spherical colloids I: Van der Waals theory}

Van der Waals
theory~\cite{Dickinson80,KinMorLin83,BelXuBau00,BelBauXu01} is the
simplest model for the liquid-gas transition, and as such is
appropriate for investigating coexistence between gas- and liquid-like
phases of colloidal suspensions (in which the structural arrangement
of the colloidal particles---for now assumed to be spherical---is
analogous to that of the atoms in ordinary gases and liquids). For
monodisperse particles, the excess free energy of van der Waals theory
is
\[
\fexc = -T\rn\ln(1-b\rn) - \half a\rn^2
\]
Here the first term represents excluded volume interactions, \ie\ the
strong short-range repulsions between colloid particles at and near
contact, with the parameter $b$ of the order of the volume of a single
particle. The second term, on the other hand, arises from
longer-ranged attractive forces between particles and is of the order
of the typical attraction energy times an interaction volume (the
latter being again of the order of the particle volume).

In the polydisperse case, $b\rn$ is generalized to $\intsig
b(\sig)\rhosig$ and $a\rn^2$ to $\int\!d\sig\,d\sig'
a(\sig,\sig')\rhosig\rh(\sig')$; the polydisperse attribute $\sig$ may
represent, for example, the diameter or charge of the colloid
particles. The functions $a(\sig,\sig')$ and $b(\sig)$ can be written
in a more physically transparent way as
\[
a(\sig,\sig') = \eps(\sig,\sig')d^3(\sig,\sig'), \qquad
b(\sig) = d^3(\sig,\sig)
\]
where $\eps(\sig,\sig')$ is the energy scale of attractions between
particles with diameter (or charge etc) $\sig$ and $\sig'$, and
$d(\sig,\sig')$ is the corresponding interaction length scale. These
functions each depend on two arguments, but can be reduced to
functions of a single argument if one assumes the so-called mixing
rules
\be
\eps(\sig,\sig') = \eps^{1/2}(\sig,\sig)\eps^{1/2}(\sig',\sig'),
\qquad
d(\sig,\sig') = \half[d(\sig,\sig)+d(\sig',\sig')]
\label{mixing_rules}
\ee
Abbreviating $\eps(\sig,\sig)$ to $\eps(\sig)$ and similarly for
$d(\sig,\sig)$, the excess free energy of polydisperse van der Waals
theory is then written as
\bea
\fexc &=& -\rn\ln\left(1-\intsig d^3(\sig)\rhosig\right)
\nonumber\\
& & {}-{} \half \int\!d\sig\,d\sig'\,
\eps^{1/2}(\sig)\eps^{1/2}(\sig') \left(\frac{d(\sig)+d(\sig')}{2}\right)^2
\rhosig\rh(\sig')
\label{vdW}
\eea
and is seen to have a truncatable structure, depending---for the most
general choice of $\eps(\sig)$ and $d(\sig)$---on six moment
densities. These have weight functions $1$, $d(\sig)$, and
$\eps^{1/2}(\sig) d^n(\sig)$ ($n=0, \ldots 3$).

Dickinson~\cite{Dickinson80} appears to have been the first to analyse
the above model. He used binning into pseudo-components to obtain
numerically some results for the ratio of the density distributions in
coexisting gas and liquid phases, which indicate the strength of
fractionation effects. He also suggested that
polydispersity might induce the qualitatively new feature of
liquid-liquid demixing, but supposed that deviations from the simple
mixing rules\eq{mixing_rules} are required for this to occur.

Gualtieri {\em et al}~\cite{GuaKinMor82} also studied the van der
Waals model for simple choices of the $\sig$-dependences, using \eg\ a
$b(\sig)$ that was constant or linear in $\sig$, together with
$a(\sig,\sig')=$ constant or $a(\sig,\sig')\propto\sig\sig'$. As
explained in Sec.~\ref{subsec:perturb}, they used a perturbation
theory approach to study the effects of the addition of a small amount
of polydisperse material to an otherwise monodisperse system,
obtaining the density distributions in coexisting phases and the
polydispersity-induced shift in the critical point. For a Schulz
parent distribution ($\rhonsig \sim \sig^\alpha e^{-\sig/\sig_0}$)
they also found the full cloud and shadow curves.

Kincaid {\em et al}~\cite{KinMorLin83} also expanded perturbatively,
but using the width $s$ of the parent distribution as the small
parameter and focussing mainly on the shift in the critical point.

Recently the van der Waals model has been revisited, for parent
distributions of Schulz or log-normal form and with various
simple choices for the functions $d(\sig)$ and
$\eps(\sig)$~\cite{BelXuBau00}. Cloud and shadow curves were found
numerically and showed small but observable changes compared to the
monodisperse case for polydispersities $s$ of the order of 10\%.  For
$s\approx 30$\% and above new critical points appear, although their
thermodynamic stability was not investigated.  Further work along
these lines~\cite{BelBauXu01} also showed that for sufficiently wide
(log-normal) parent distributions three-phase coexistence can occur,
even for the simple mixing rules\eq{mixing_rules} above. In fact one
can say rather more: If, as in Ref.~\cite{BelBauXu01}, one assumes
$d(\sig)=$ constant---so that the only effect of polydispersity is on
the attraction energy parameters $\eps(\sig)$---then in the dense
limit the model formally maps to the Flory-Huggins theory of a random
copolymer blend~\cite{Sollich_colpol} discussed in
Sec.~\ref{sec:copol}. 
It can therefore show liquid-liquid demixing into an {\em
arbitrarily large} number of phases as temperature is lowered, and can
also exhibit critical points of arbitrarily high order.

\subsection{Spherical colloids II: Hard spheres}

Van der Waals theory does not address the question of crystallization
in colloidal suspensions, where the particles arrange themselves into
a lattice structure with long-range translational 
order. The ``cleanest'' system for
studying this transition is one where the colloidal particles act as
{\em hard spheres}, exhibiting no interaction except for an infinite
repulsion on overlap. This scenario can indeed be realized experimentally,
using for example latex particles that are sterically stabilized by a
polymer coating~\cite{PusVan86}. In a hard sphere system the only
energy scale is set by the temperature; $T$ therefore only appears as
a trivial scaling factor in the 
results and will be set to unity in this section. There is also
no gas-liquid transition, so it is common to refer to the
non-crystalline phase of hard spheres as a
fluid (rather than a gas or a liquid).
Monodisperse hard spheres exhibit only a freezing
transition, where a fluid with a volume fraction $\phi$ of spheres of
$\phi\approx 50$\% coexists with a crystalline solid with $\phi\approx
55$\%. Phase separation is observed when the overall volume fraction
of the system lies between the values for the coexisting fluid and
crystal; for $\phi<50$\%, on the other hand, 
one has only the fluid and for $\phi>55\%$
(and up to the maximum close-packed value of $\phi\approx 74$\%) only
the solid.

For colloidal hard spheres, there is inevitably some polydispersity in
the diameter $\sig$ of the spheres. It was realized early on that such
diameter polydispersity might destabilize the colloidal crystal phase,
eventually inhibiting freezing above a certain ``terminal''
polydispersity. Experimentally, the freezing transition is indeed
suppressed in sufficiently polydisperse
systems~\cite{PusVan86,Pusey91}. But the situation is somewhat
ambiguous, since the observed terminal polydispersity might also be a
non-equilibrium effect due to a kinetic glass
transition~\cite{PusVan87}; the growth kinetics of polydisperse
crystals may also cause deviations from equilibrium
behaviour~\cite{EvaHol01}. The determination of an accurate {\em
equilibrium} phase diagram for polydisperse hard spheres is
nevertheless an important task, if only to allow experimental findings
to be properly attributed to equilibrium or non-equilibrium
effects. The results could also guide future experiments on colloidal
suspensions under microgravity conditions, where---with the glass
transition shifted to higher densities or even
absent~\cite{ZhuLiRogMeyOttRusCha97}---more of the equilibrium
behaviour should be observable. In the remainder of this section, and
in keeping with the overall focus of the paper, I will therefore focus
on attempts to clarify the {\em equilibrium} phase behaviour of
polydisperse hard spheres.

Much early theoretical work (see~\cite{PhaRusZhuCha98} for a
comprehensive list of references) focussed on estimating the terminal
value $\st$ of the polydispersity $s$. As above, $s$ is defined as the
normalized standard deviation of the diameter distribution; see
eq.\eq{s_def}. Dickinson {\em et al}, for example, extrapolated the
decrease of the volume change on melting with polydispersity $s$ to
zero, obtaining an estimate of $\st\approx
30$\%~\cite{DicPar85}. Pusey~\cite{Pusey87} used a simple
Lindemann-type criterion to estimate that the larger spheres in a
polydisperse system would disrupt the crystal structure above
$\st\approx 6\ldots12$\%. McRae and Haymet~\cite{McrHay88} used
density functional theory (DFT---see Sec.~\ref{sec:inhomogen})
together with the simplifying
assumption that there is no fractionation, \ie\ that fluid and crystal
have the same distribution of diameters, and found that there was no
crystallization above $\st\approx 5\%$. Barrat and
Hansen~\cite{BarHan86} also employed DFT, estimating the free energy
difference between fluid and solid; while in the monodisperse case the
solid has the lower free energy above volume fraction $\phi=55$\%, the
fluid can become preferred again at large $\phi$ if the polydispersity
$s$ is sufficiently large. This result is compatible with the
intuition that polydispersity {\em reduces} the maximum packing
fraction in a crystal (since a range of diameters need to be
accommodated on uniformly spaced lattice sites), while it {\em
increases} the maximum packing fraction in the fluid, where smaller
spheres should be able to fill ``holes'' between larger particles more
easily. A more detailed calculation~\cite{PhaRusZhuCha98} confirmed
this, estimating the terminal polydispersity from the crossing of the
maximum packing fractions of liquid and solid as $\st=12\%$.

In recent years, computer simulations have also been used to estimate
the terminal polydispersity. It is difficult, however, to carry out
such simulations for the experimentally most relevant situation of a
fixed parental density distribution $\rhonsig$: With a number of
particles that can be simulated of the order of hundreds, there will
be strong finite size effects due to the random assignment of
diameters to particles; furthermore, with only a few particles in each
small range of diameters, it is almost impossible to ensure that
the size distributions in coexisting phases are properly
equilibrated. Instead, a semi-grandcanonical approach has been used,
which prescribes the differences in chemical potential $\mu(\sig)$
between different $\sig$; effectively, one then simulates a system
with variable polydispersity. Bolhuis and Kofke, for example, imposed
a parabolic shape for the chemical potential differences, giving a
Gaussian distribution of diameters at low density~\cite{BolKof96}.
Using thermodynamic integration they then followed the pressure at
which fluid-solid coexistence occurs as a function of the width of
this Gaussian distribution. They found that this coexistence line
terminates, at a point where the densest packings for fluid and solid
were reached; the diameter distributions there were significantly
different, with the fluid having a polydispersity of
$s\approx 12\%$ and the solid
$s\approx 6\%$. However, this terminal point is of limited relevance,
since it only exists {\em for the given chemical potential
differences}. One can in fact go beyond it by considering more general
functional forms for the chemical potential
differences~\cite{KofBol99}; nevertheless, Kofke and Bolhuis observed that
the coexisting solid always seemed to have a
polydispersity below
$s\approx 6\%$, while for the fluid much larger values of $s$ could be reached.
(An unpublished preprint by Almarza and
Enciso~\cite{AlmEnc99} comes to similar conclusions.) Based on this
observation, it was suggested~\cite{KofBol99} that a polydisperse hard
sphere fluid may freeze by splitting off a series of solids comprising
a narrow range of (large) sphere diameters each\footnote{%
In a more extreme scenario, where the diameter distribution has a fat
(slower than exponential) tail extending to very large values,
Sear~\protect\cite{Sear99c} argued that such a
fractionated solid would in fact occur already at vanishingly small densities.
}.

While the simulation results described above are suggestive, they are
still obtained for variable polydispersity, \ie\ by fixing chemical
potential differences. In contrast to the experimental situation, the
overall particle size distribution can thus change (sometimes
dramatically) across the phase diagram, limiting the applicability of
the results\footnote{%
A simulation technique to address this problem is currently being
developed~\cite{WilSol02}.}%
. A number of researchers have therefore tried to investigate the
phase behaviour of polydisperse hard spheres theoretically, using
approximate expressions for the (excess) free energy. For the fluid
phase, the most accurate such approximation available is currently
believed to be the generalization by Salacuse and
Stell~\cite{SalSte82} of the BMCSL equation of
state~\cite{Boublik70,ManCarStaLel71}); for the monodisperse case this
reproduces the well-known Carnahan-Starling equation of
state. Assuming that sphere diameters are measured in units of some
reference value $\sig_0$, and that all densities are made
non-dimensional by multiplying by the volume $\pi\sig_0^3/6$ of a
reference sphere, the BMCSL expression for the excess free energy
is
\be
\fexc = \left(\frac{\rho_2^3}{\rho_3^2} - \rho_0\right)\ln(1-\rho_3) +
\frac{3\rho_1\rho_2}{1-\rho_3} + \frac{\rho_2^3}{\rho_3(1-\rho_3)^2}
\ee
This has again a truncatable form, involving only the (ordinary)
moments $\rho_i=\intsig \sig^i \rhosig$ ($i=0\ldots 3$) of the density
distribution; with our choice of units $\rho_3\equiv \phi$ is the
volume fraction of spheres. Bartlett~\cite{Bartlett99} provided an
elegant argument why---at least within a virial expansion---such a
moment structure of the excess free energy for the hard sphere fluid 
should in fact be exact.

For phase coexistence calculations it is desirable also to have a
compact expression for the excess free energy of the polydisperse hard
sphere {\em crystal}. This is not at all a trivial question, in particular
since the structure of such a crystal could be rather complex, with
different sites inside the crystalline unit cell occupied
preferentially by particles with different ranges of diameters. Most
theoretical work therefore assumes that one instead has a
substitutional solid, where crystal sites are assumed to be occupied
equally likely by particles of any diameter. A simple-minded but
popular approach to estimating the free energy is then cell theory,
where particles are treated as independent but confined in an
effective cell formed by their neighbours (see \eg~\cite{Sear98}). A
more quantitative, ``geometric'' approach has recently been proposed
by Bartlett~\cite{Bartlett99,Bartlett97}: He assumed that the excess
free energy of the solid depends on the same \moms\ $\rn$, \ldots,
$\rh_3$ as that of the fluid, and then fitted the functional form of this
dependence 	by comparing with 
simulation data on bidisperse hard sphere systems.

\begin{figure}
\begin{center}
\epsfig{file=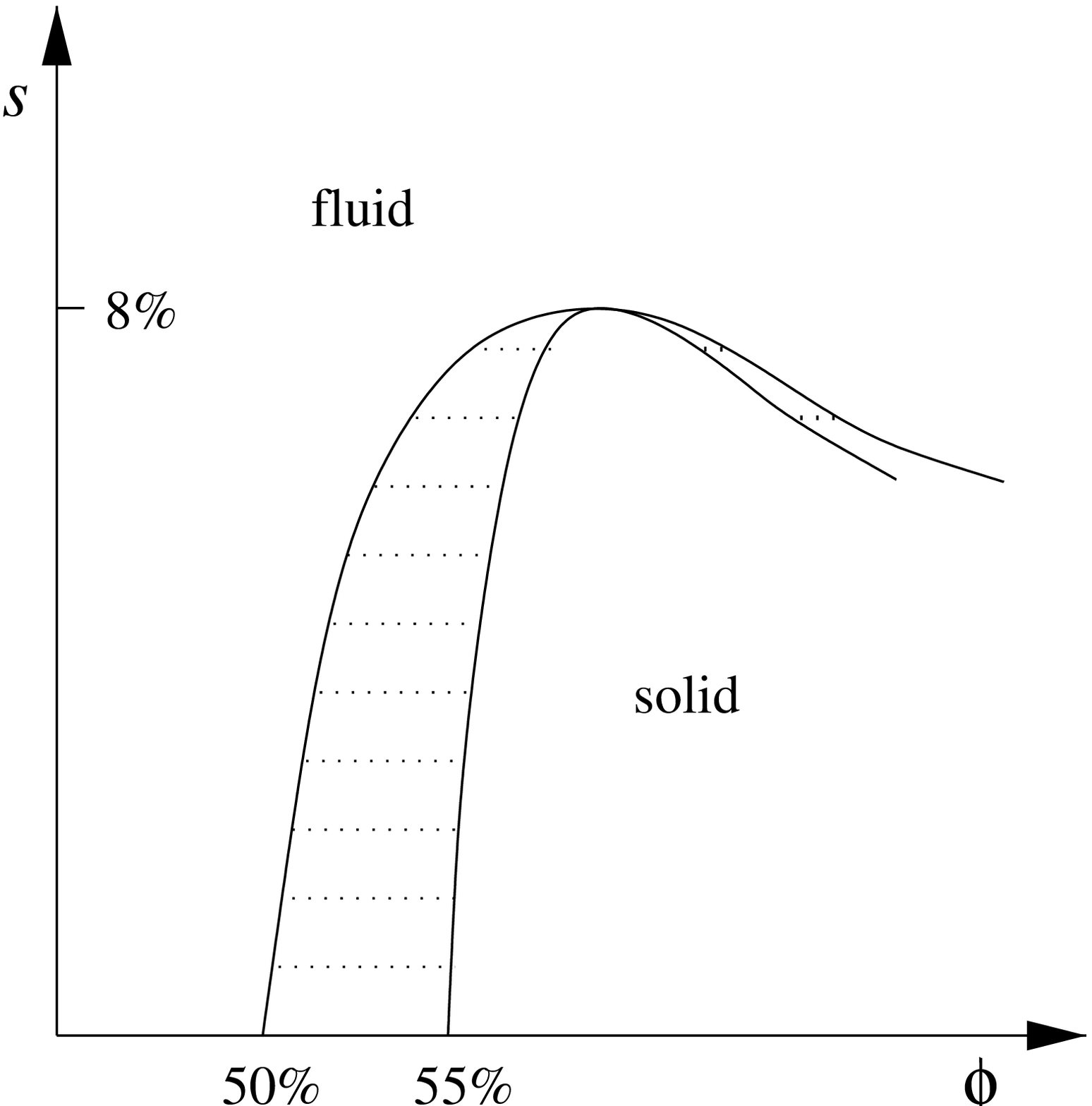,width=0.4\textwidth}\hspace*{5mm}%
\epsfig{file=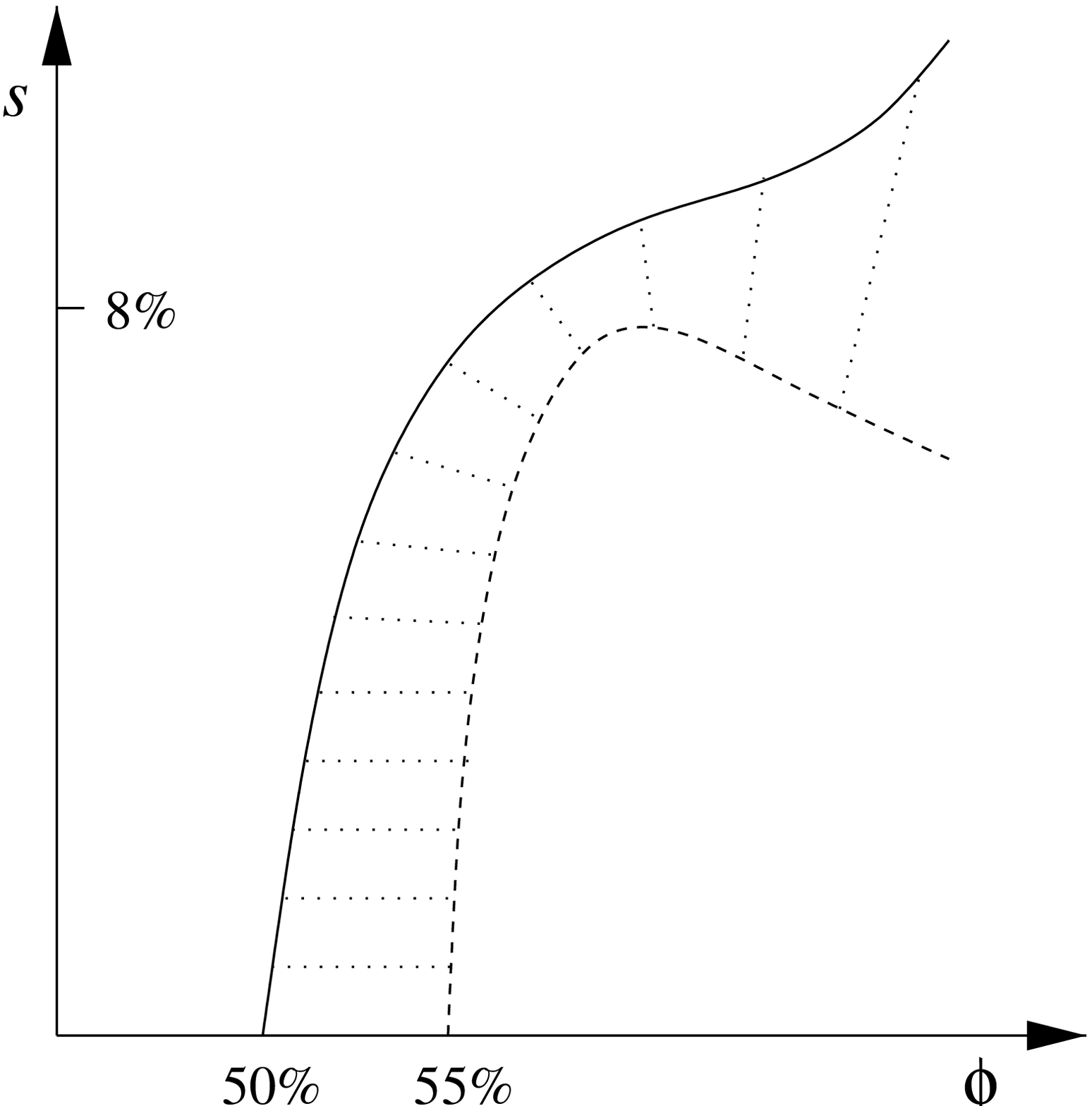,width=0.4\textwidth}
\end{center}
\caption{Left: Sketch of Bartlett and Warren's phase diagram for
polydisperse hard spheres~\protect\cite{BarWar99}. Shown is the result
of their simplest approximation, in which fractionation is not
allowed; cloud and shadow curves then effectively reduce to the 
conventional coexistence curves for monodisperse systems,
and coexisting phases (connected by dotted lines) share the same value
of the polydispersity $s$. (A better approximation used
in Ref.~\protect\cite{BarWar99} allowed
fractionation but still effectively constrained $s$ to be equal in
coexisting phases.) Note the re-entrant transition to a fluid at large
colloid volume fraction $\phi$, for polydispersities $s$ just below
the terminal value $\st\approx 8\%$. At $\st$, the phase boundaries
meet in a ``point of equal concentration''. Right: Possible shape of
the ``true'' phase diagram that would result from a calculation which
allows for different polydispersities of fluid and solid. Shown are a
cloud curve from the fluid side, and the corresponding curve of solid
shadow phases; coexisting phases are again connected by dotted
lines. There is now no reason why cloud and shadow curves should meet, so
that a point of equal concentration seems unlikely. There may also not
be a terminal value of the polydispersity on the fluid cloud curve,
since even a very polydisperse fluid may 
always be able to split off a solid with a narrow
distribution of particle diameters. At large values of $s$,
fluid-fluid demixing (not shown) might pre-empt the fluid-solid transition.
\label{fig:HS_reentr}
}
\end{figure}

By applying the moment free energy method to the BMCSL free energy for
the fluid and Bartlett's ``geometric'' free energy for the solid,
Bartlett and Warren~\cite{BarWar99} recently investigated the freezing
behaviour of polydisperse hard spheres. They found that the range of
volume fractions where fluid-solid coexistence is observed narrows as
the polydispersity $s$ is increased and eventually shrinks to zero at
a terminal polydispersity $\st\approx 8\%$. (At this point, the
density distributions in the fluid and solid were calculated to be
equal, but as the symmetries of the two phases
are different this is not a critical
point but rather a ``point of equal concentration''~\cite{BarWar99}.)
At values of $s$ just below $\st$, they also found a transition to a
re-entrant fluid at large volume fractions (see
Fig.~\ref{fig:HS_reentr}(left)). Such a re-entrance is in fact to be
expected from the earlier work on the terminal polydispersity
described above: For sufficiently polydisperse systems, the fluid
should at large volume fractions be thermodynamically preferred over
the solid because it packs the spheres more efficiently.  When
interpreting the results of~\cite{BarWar99}, however, it needs to be
born in mind that the approximations made in effect constrained the
polydispersity (normalized standard deviation) $s$ to be equal in the
coexisting fluids and solids, allowing only the mean sizes to be
different. The possibility of a very polydisperse fluid splitting off
a solid containing a narrow range of diameters is thus disallowed.
Work is in progress to remove these
simplifications~\cite{Sollich_colpol}, and one may speculate that the
point of equal concentration would disappear in a more accurate
treatment (see Fig.~\ref{fig:HS_reentr}(right)).

The analysis of the freezing behaviour of strongly polydisperse
hard spheres is complicated by the fact that, instead of a single
solid phase, a number of coexisting solids with strong diameter
fractionation between them may appear. Bartlett~\cite{Bartlett98} and
Sear~\cite{Sear98} both investigated this possibility, using different
approximations for the free energy for the solid, and found that an
increasing number of fractionated solids should appear as the system
is made more polydisperse. Both calculations only compared the free
energies of the liquid and the fractionated solids, however, rather
than solving the full phase equilibrium conditions. They also used the
drastic assumption that the different solids would split the range of
diameters evenly between themselves, so that spheres of 
any given diameter would
occur in only a single phase; in reality, one would expect a rather
more gradual fractionation of the phases.

One final complication in the phase behaviour of strongly polydisperse
hard sphere fluids is the possibility of fluid-fluid demixing. While
for {\em bidisperse} hard spheres such a demixing transition is believed to
be absent (or at least always metastable compared to the freezing
transition) Warren~\cite{Warren99} found, using the BMCSL free energy,
that for a bimodal diameter distribution a demixing instability could
occur at reasonable volume fractions. Warren noticed that significant
polydispersity ($s\geq 50\%$) in the larger spheres was necessary to
produce this effect. He conjectured that the demixing occurred because
the smaller spheres cause an effective attraction (depletion
interaction, see Sec.~\ref{sec:colloid_polymer} below)
between the larger spheres; polydispersity
then facilitates the demixing by making a dense phase of larger
spheres more favourable (due to the increased maximal packing
fraction). Cuesta~\cite{Cuesta99} studied log-normal diameter
distributions; even though these only have a single maximum, and thus
no separation into small and large spheres akin to the bimodal case,
he still predicted fluid-fluid demixing for large polydispersities
($s\geq 160\%$).

The theoretical studies reviewed above still leave open a substantial
number of questions. For fluid-fluid demixing, for example, only the
spinodal instability was analysed~\cite{Cuesta99,Warren99}. The actual
demixing transition will occur at a lower density yet to be
determined; and no predictions exist for the freezing behaviour of
such demixed fluids at higher densities. The drastic---and
differing---approximations for size fractionation that were used in
the studies of re-entrant melting and solid-solid
coexistence~\cite{Sear98,BarWar99,Bartlett98} also leave the relative
importance of these two phenomena unclear.  Work is now underway to
address these questions and produce a coherent picture of the
equilibrium phase behaviour of polydisperse hard spheres~\cite{Sollich_colpol}.

\subsection{Colloid-polymer mixtures}
\label{sec:colloid_polymer}

Moving beyond suspensions of (hard) spherical colloids alone,
colloid-polymer mixtures have in recent years attracted considerable
interest, mainly because the polymer induces an easily tunable
``depletion interaction'' between the colloids. This
interaction arises as follows. When colloidal particles approach each
other to within twice the radius of gyration (\ie\ the effective
diameter) of the polymer chains, they form a ``depletion zone''
between them which the polymer chains cannot enter. The result is an
imbalance in the polymer osmotic pressure which pushes the colloidal
particles together, causing an effective colloid-colloid
attraction. This attraction can lead to the appearance of a
(colloidal) gas-liquid coexistence region in the phase
diagram~\cite{IleOrrPooPus95}; its range and strength are tunable
via the size of the polymer chains and the polymer concentration,
respectively. This feature makes colloid-polymer mixtures interesting
model systems with which to study the conditions required for the
appearance of liquid phases; theory, simulation and experiment all
reveal, for example, that the interaction range needs to exceed a
certain fraction of the particle size (of order $30\%$) in order for
gas-liquid coexistence be stable rather than
metastable~\cite{IleOrrPooPus95,LekPooPusStrWar92,DijBraEva99}.

To model the simplest case of colloids with hard interactions and
ideal polymers (in a so-called $\theta$-solvent), the Asakura-Oosawa
model~\cite{AsaOos54} has been widely used. It treats the polymer
coils as spherical particles that can interpenetrate freely with each
other, but experience a hard sphere repulsion when they come into
contact with the colloids. Formally integrating out the polymer
degrees of freedom then results in the expected attractive
colloid-colloid interaction. However, this interaction generally
contains many-body terms (arising from the overlap of the depletion
zones of more than two colloids)~\cite{DijBraEva99} and so its effect
is difficult to take into account exactly. But progress can be
made using a van der Waals (mean field) type of approach which replaces
the effective colloid-colloid interaction by its average over the pure
(hard sphere) colloid system~\cite{LekPooPusStrWar92}. In the case of
monodisperse colloid, the resulting phase behaviour is well
understood, with the main feature being the appearance of gas-liquid
phase separation; non-ideality of the polymer chains can also be
included in the model but only introduces a weak temperature
dependence into the phase behaviour~\cite{WarIlePoo95}. Polydispersity
in polymer chain lengths has been studied~\cite{SeaFre97,LeeRob99},
but only for variable polydispersity where the chemical potential
differences between chains of different length are imposed. 
Warren~\cite{Warren97} considered instead the
experimentally more relevant situation where the overall
polymer density distribution is imposed, in the
simpler case where the polymer consists of a binary mixture of chains
of two different lengths. He made the intriguing
observation that polydispersity has almost no effect on the phase
behaviour as long as the polymer concentration is expressed in terms
of an effective volume fraction (which allocates to each polymer chain
a volume proportional to the cube of its radius of gyration). The
generalization to a fully polydisperse polymer with imposed density
distribution is challenging, but work in this direction is in
progress~\cite{Sollich_colpol}.

The results reviewed above all concern the case of colloidal particles
of identical size. For the more complicated case of polydisperse
colloids, only rough qualitative estimates of the effects on phase
behaviour~\cite{Fairhurst99} and limited perturbative results for
narrow size distributions~\cite{Evans01} exist. Recent experimental
results~\cite{Fairhurst99} do, however, suggest that for fully
polydisperse colloids intricate---and largely unexplored---phase
diagram topologies may occur, due to the combination of gas-liquid
coexistence on the one hand and re-entrant melting in the absence of
polymer on the other. The theoretical analysis of these effects
remains an open problem, but should be helped by the fact that, even
for the most general case of polydisperse colloid diameters {\em and}
polydisperse polymer chain lengths, the van der Waals treatment
of Ref.~\cite{LekPooPusStrWar92} leads to a truncatable structure for the
free energy~\cite{Sollich_colpol}.

\subsection{Colloidal liquid crystals I: Maier-Saupe theory for
thermotropics}

So far I have only discussed spherical colloids. Non-spherical
particles, shaped \eg\ like rods or plates, can form liquid
crystalline phases; these are the subject of the following sections.
One of the simplest liquid crystal structures is the nematic. Like a
liquid, it has no long-range translational order, but the rods are
{\em orientationally} ordered, pointing preferentially along a fixed
direction called the nematic axis. The density distribution required
to describe a nematic phase of length-polydisperse rod-like particles
thus depends on two variables, the rod length and the rod orientation.
Since the orientation of a rod can and will change over time, one has
a mixture of conserved and non-conserved degrees of freedom, and this
makes the problem rather challenging.

Liquid crystals in which phase transitions are driven primarily by
changes in temperature (rather than density) are called {\em
thermotropic}. The standard model for analysing their phase behaviour
is Maier-Saupe theory~\cite{MaiSau58}, which captures the
orientation-dependent attractions between particles. It was originally
derived on the basis of an approximate treatment of the van der Waals
attraction between large molecules, caused by fluctuating charge
densities in their electron clouds, but is actually much more widely
applicable as a phenomenological theory of orientation-dependent
interparticle attractions.

Consistent with the physical intuition that in thermotropics phase
transition are driven by temperature variations rather than changes in
density, Maier-Saupe theory effectively neglects changes in the
overall particle density, so that different phases only differ in
their normalized distributions $\prob(L,\Omega)$ over rod lengths $L$
and orientations $\Omega$. With the density having been fixed, it is
sensible to switch from the free energy density $f=F/V$ to the free
energy per particle $F/N$ as the basic quantity from which to analyse
phase behaviour; the non-ideal part of this is, for Maier-Saupe theory
\be
\frac{\tilde{F}}{N} = -\half
\int\!\!dL\,dL'\,d\Omega\,d\Omega'\,
\prob(L,\Omega)\prob(L',\Omega) u(L,L') P_2(\cos\theta)P_2(\cos\theta')
\label{Maier_Saupe}
\ee
The main ingredient of this expression is the angular dependence
through the second-order Legendre polynomials
$P_2(\cos\theta)=(3\cos^2\theta-1)/2$; here $\theta$ is the angle of a
rod with the nematic axis. The excess free energy\eq{Maier_Saupe}
favours nematic ordering, as it would be minimal if all rods pointed
along the nematic axis ($\theta=0$). The ideal part $T\int\!\!
dL\,d\Omega \,\prob(L,\Omega)[\ln\prob(L,\Omega)-1]$ of the free energy
per particle instead prefers an isotropic phase (which, due to its
random rod orientations, has the largest orientational entropy).

In the monodisperse case, where there is only a single rod length $L$,
Maier-Saupe theory leads to a transition from an isotropic to a nematic as
the temperature is lowered. This is consistent with the intuition
explained above; the scale for the transition temperature is set by
the energy scale for the attractive interaction, $u(L,L)$. Note that
there is no coexistence gap here, \ie\ no temperature region here
where isotropic-nematic (I-N) phase coexistence is observed. This is
because the only conserved density is the total particle density,
which is assumed equal in all phases.

In the polydisperse case, the (essentially phenomenological) function
$u(L,L')$ determines how the strength of the attraction varies with
the rod lengths. Now there are nontrivial conserved densities: The
length distribution $\prob\pn(L)$ of the parent phase has to be
maintained, and the system may be able to lower its free energy by separating
into two phases with different length distributions. Accordingly,
Sluckin~\cite{Sluckin89} found in a perturbative calculation for
narrow polydispersity that a coexistence gap develops in the
polydisperse system; the temperature range over which I-N coexistence
is observed is proportional to the variance $s^2$ of the parent length
distribution. Since the function $u(L,L')$ is of a phenomenological nature, the
same conclusion also applies if the polydisperse attribute is
different, \eg\ rod diameter or charge instead of length.

The effects of stronger polydispersity (which cannot be treated
perturbatively) in the Maier-Saupe model remains unexplored;
one interesting question that could be asked~\cite{SpeSol} is whether
coexistence between several nematic phases would eventually develop,
as it does in the Onsager model discussed next.

\subsection{Colloidal liquid crystals II: Isotropic-nematic
transition in hard rods}

In {\em lyotropic} liquid crystals, the important control parameter
causing phase transitions is density, rather than temperature as in
thermotropics. The paradigmatic model for lyotropic colloidal liquid
crystals is that of Onsager~\cite{Onsager49}, which neglects any
long-range attractions between particles and only retains the
short-range repulsions; the latter are taken to be hard (infinite
repulsion on contact). Rod-like colloidal particles approximating
closely such ``hard rods'' can be realized experimentally
(see \eg~\cite{BuiLek93}). Because of the hard interactions in the
Onsager model, the temperature can be trivially scaled out of all
results and will be set to unity below.

Onsager's treatment of the hard rod model is based on a virial
expansion in the overall particle density. Crucially, it turns out
that for long thin rods, and in the region of densities where the I-N
phase transition occurs, this virial expansion can be {\em exactly}
truncated after the first nontrivial (second virial) contribution. The
intuitive reason for this is as follows. Assume the rods have
cylindrical shape, with length $L$ and diameter $D$. Any given rod
excludes another, randomly oriented rod from a volume of $\order(L^2
D)$. The I-N transition occurs at densities $\rn$ where the number of rods
in this excluded volume becomes of order one, giving $\rn\sim
L^{-2}D^{-1}$. Multiplying by the rod volume ($\sim LD^2$) gives the
rod volume fraction $\phi\sim D/L$ at the transition. For long thin
rods this becomes vanishingly small, making it plausible that 
higher order terms in the virial expansion can be neglected.

To state the free energy of a system of long thin rods with
polydisperse lengths $L$ and diameters $D$, let us choose a reference
length $L_0$ and reference diameter $D_0$, and define normalized
lengths $\Ln=L/L_0$ and diameters $\Dn=D/D_0$. It is conventional to
choose $(\pi/4)L_0^2 D_0$ as the unit volume to make
densities non-dimensional; this is the
average excluded volume of two randomly oriented reference rods.  In
the Onsager limit $D_0/L_0\to 0$ (at fixed distribution of $\Ln$ and
$\Dn$), the excess free energy of this hard rod system then
becomes~\cite{VroLek92}
\be
\fl
\fexc = \frac{4}{\pi}
\int\!\!d\Ln\,d\Ln'\,d\Dn\,d\Dn'\,d\Omega\,d\Omega'\,
\rho(\Ln,\Dn,\Omega)\rho(\Ln',\Dn',\Omega') \,\Ln\Ln'\,\frac{\Dn+\Dn'}{2}\,
|\sin\gamma(\Omega,\Omega')|
\label{Onsager}
\ee
The nontrivial factors in this expression arise from the fact that the
excluded volume of two rods making an angle $\gamma$ with each other
is $LL'(D+D')|\sin\gamma|$, or $(4/\pi)\Ln\Ln'(\Dn+\Dn')|\sin\gamma|$ in our
volume units. The excess free energy is a functional of the density
distribution $\rho(\Ln,\Dn,\Omega)$, which is defined such that
$\rho(\Ln,\Dn,\Omega)d\Ln\,d\Dn\,d\Omega$ is the density of rods with
lengths in an interval $d\Ln$ around $\Ln$, diameters in an interval $d\Dn$
around $\Dn$, and orientations in a solid angle $d\Omega$ around
$\Omega$.

If the orientation $\Omega$ is parameterized in terms of the rod angle
$\theta$ with the nematic axis, and the azimuthal angle $\varphi$,
then the density distributions are independent of $\varphi$ and the
integrations over $\varphi$ and $\varphi'$ in\eq{Onsager} can be
carried out, defining a function
\be
K(\theta,\theta') = \frac{4}{\pi} \int\! d\varphi\,
d\varphi'\, |\sin\gamma(\Omega,\Omega')|
\label{K}
\ee
which encodes the angular dependence of the excluded volume interaction.

For the case of monodisperse rods (see Ref.~\cite{VroLek92} for a
comprehensive review), one sets $\Dn=\Dn'=1$ and $\Ln=\Ln'=1$ everywhere
in\eq{Onsager}; the density distribution then becomes a function
$\rh(\theta)$ of only the rod angle $\theta$ with the nematic
axis. One can separate this into its conserved and non-conserved parts
by writing $\rh(\theta)=\rn\prob(\theta)$; the normalized orientation
distribution function $\prob(\theta)$ 
is found for any given $\rn$ by minimizing the free energy.
This gives, at least conceptually, the free energy as a function of
$\rn$; a double tangent construction then shows a coexistence gap,
across which an isotropic phase of density $\rho_0\approx 3.29$ coexists with
a nematic phase with $\rho_0\approx 4.19$.

Consider now the case of length polydispersity (with the diameters
still monodisperse). Previous work in this area has focussed on the
simplified case of bi- and tridisperse mixtures (rods with two or
three different lengths) and has uncovered---for sufficiently
disparate lengths---a number of features not observed in the
monodisperse case. These include the possibility of coexistence of
several nematic phases (N-N), possibly also together with an isotropic
phase
(I-N-N)~\cite{LekCouVanDeb84,OdiLek85,VroLek93,VanMul96,Hemmer99};
such an I-N-N coexistence has indeed been observed
experimentally~\cite{BuiLek93}. In the tridisperse
case~\cite{VroLek97}, up to four phases (I-N-N-N) can coexist. For
bidisperse systems with length ratios above $\approx 5$, re-entrant
phase coexistence sequences such as I $\to$ I-N $\to$ N $\to$ I-N
$\to$ N are also found~\cite{LekCouVanDeb84}. At rod volume fractions far
above the onset of I-N coexistence (but, due to the Onsager limit
$D_0/L_0\to 0$, still negligible compared to unity), the phase diagram
is predicted to be density independent, so that N-N coexistences
persist rather than being terminated by a critical point at high
density~\cite{VroLek93,VanMul96,VroLek97}.
For bidisperse {\em diameters} and monodisperse lengths, I-I
demixing~\cite{SeaJac95,VanMulDij98} and I-I-N coexistence can occur
as additional features; a nice discussion on why multiple
isotropic phases require diameter polydispersity can be found in
Ref.~\cite{VanMulDij98}.

The studies described above show that a wealth of new phase behaviour
can result even for bidisperse hard rod systems. For the potentially
even richer case of true polydispersity, however, results to
date are very limited. The only studies of the full Onsager model are
perturbative calculations, which show a widening of the
coexistence gap at the I-N transition~\cite{Sluckin89,Chen94} with
increasing length polydispersity\footnote{%
Chen's analysis~\cite{Chen94} provides an instructive example of the
importance of taking fractionation into account when studying
polydisperse phase behaviour. The coexistence region for a parent
phase with a given length distribution is bounded by
the isotropic cloud point---where the nematic phase first
appears---at the lower end, and the nematic cloud point---where the
fractional volume occupied by the isotropic phase goes to zero---at
the upper end. Chen instead found the densities of the isotropic cloud
phase and its coexisting shadow. The gap between these two densities
{\em decreases} as polydispersity increases, while the width of the
coexistence region {\em increases}.
}%
. For the simpler Zwanzig model of rods oriented along one of three
perpendicular axes, a full treatment of the length polydisperse
case~\cite{ClaCueSeaSolSpe00} has recently confirmed this trend.
However, no evidence of N-N coexistence was found, even for
significant polydispersities; an earlier calculation for the
bidisperse gave similar results~\cite{ClaMcl92}. This contrast to the
predictions of the full Onsager model can be explained intuitively as
follows: When a polydisperse nematic phase splits into two nematics
containing predominantly short and long rods, respectively, it gives
up entropy of mixing but gains orientational entropy. In the Onsager
model, where the rod angles are continuous variables, the gain in
orientational entropy can be arbitrarily large, thus favouring such a
phase split. (The orientational entropy tends to $-\infty$ as the
orientational distribution function tends to a delta function.)  In
the Zwanzig case, on the other hand, the maximum gain in orientational
entropy is $k_{\rm B}\ln 3$ (this being the difference between the
entropies of an isotropic and a fully ordered nematic phase) so that
nematic-nematic coexistence is disfavoured.

It is clear, then, that a number of open questions remain regarding
the effects of polydispersity in the Onsager model of hard rods. In
particular, one would like to know under which conditions on the width
and/or shape of the length and diameter distributions N-N, I-N-N and
I-I phase coexistences are possible. The answers cannot be inferred
from the results for the bi- or tridisperse cases; otherwise one
would incorrectly predict, for example, that any polydisperse system
should show N-N coexistence since it contains {\em some} rods of very
different lengths. Equally, it remains unclear how many nematic
phases can coexist far above the I-N transition, where the phase
diagram becomes density-independent. Genuine polydispersity could also
cause entirely new effects, \eg\ demixing into more than two isotropic
phases for sufficiently wide diameter distributions.

Tackling the polydisperse Onsager model head on is difficult, since
the excess free energy\eq{Onsager} does not have a truncatable
structure. However, one can exploit the known expansion of the angular
part $K(\theta,\theta')$ of the excluded volume interaction
(see\eq{K}) in terms of Legendre polynomials. This takes the
form~\cite{KayRav78}
\be
K(\theta,\theta') = c_0 - \sum_{n=1}^\infty c_n
P_{2n}(\cos\theta)P_{2n}(\cos\theta')
\label{K_exp}
\ee
with positive constants $c_n$. Truncating this series at successively
higher order, one recovers
truncatable systems which approach the
full Onsager model in the limit; the moment densities that occur are
defined by the weight functions $w_n(\Ln,\Dn,\theta) = \Ln
P_{2n}(\cos\theta)$ (as well as $\Ln\Dn P_{2n}(\cos\theta)$ if 
diameter polydispersity is present). Judging from existing work on the
monodisperse case~\cite{LekCouVanDeb84}, even the lowest nontrivial
order of truncation---which, for length polydispersity, gives one
conserved and one non-conserved moment density---should already give
qualitatively correct results~\cite{SpeSol}.

\subsection{Colloidal liquid crystals III: Hard rods at higher
densities}

Rod-like colloidal particles should, at sufficiently high densities,
form crystalline solids; a smectic phase (where the particles are
arranged into layers that are perpendicular to their preferred
orientation, but lack translational order within the layers) may
also intervene between the nematic and the crystal phase. Neither
smectic nor crystal phases are accessible within the Onsager theory of
long thin rods as outlined above, however: They occur at rod volume
fractions of order unity, and hence densities $\rn \sim L^{-1}D^{-2}$;
the densities $\rn \sim L^{-2}D^{-1}$ at the isotropic-nematic
transition are much smaller (in fact infinitely so, in the
Onsager limit $D/L\to 0$).

Studying the effects of polydispersity in this high density regime is
an enormous challenge, in part because there is still significant
controversy over the most appropriate free energy functionals in this
region of the phase
diagram~\cite{BatFre98,BohHolVil96,MarCueRoiMul99}. Some qualitative
features are known, however. Significant length polydispersity, for
example, should make smectic (and possibly also crystalline) phases
less favourable, since a broad range of rod lengths will be
difficult to accommodate within these structures.
Instead, one expects to see columnar
phases, where the rods are arranged into columns which are themselves 
packed into a
two-dimensional hexagonal lattice; since the rods can slide freely
within each column, such columnar phases can easily tolerate a
spread of rod lengths. Sluckin~\cite{Sluckin89} indeed found, within a
perturbative treatment, that the onset of smectic order should be
delayed (\ie\ shifted to higher densities) by polydispersity in rod
lengths, and that eventually the smectic phase should disappear in
favour of a columnar phase.  Bates and Frenkel~\cite{BatFre98} arrived
at similar conclusions from their semi-grandcanonical (variable
polydispersity) simulations: When the polydispersity increased beyond 
a terminal value of 
$s\approx 18\%$, the smectic phase was no longer stable. They also
argued that length polydispersity should destabilize the crystal in
favour of the columnar phase, though disagreeing on the density
dependence of the relative stability of the two phases with an earlier
density functional treatment~\cite{BohHolVil96}.

One final new effect of length polydispersity on hard rod phases at
high densities is the possibility that, on increasing the density,
nematic-nematic demixing might occur before the transition to a
smectic or columnar phase~\cite{VanMul96b}. This seems entirely
plausible, given that the Onsager treatment described above predicts
N-N demixing (in sufficiently bidisperse systems) at densities
arbitrarily far above the I-N transition. The behaviour of such
demixed nematics at higher densities is an entirely open question;
they might, for example, form two demixed (fractionated) smectics
rather than a single columnar phase. Another area that remains
unexplored is the effect of {\em diameter} polydispersity on the high density
behaviour: This would be expected, for example, to disadvantage
columnar phases against smectics, thus producing an effect opposite to
that of length polydispersity.

\subsection{Colloidal liquid crystals IV: Plates and rod-plate mixtures}

Liquid crystalline phases can also occur in suspensions of {\em plate-like}
(rather than rod-like) particles. If the particles have hard
interactions and are monodisperse, then one expects the sequence
isotropic (I) $\to$ nematic (N) $\to$ columnar $\to$ crystalline as
the particle density is increased. As for rod-like particles, the I-N
transition (observed experimentally in Ref.~\cite{VanLek98}) can
be analysed using Onsager's second virial theory, although due to the
different scaling of the higher order virial coefficients the results
do not become exact even in the limit of very thin plates.

The effect of polydispersity on the phase behaviour of plate-like
colloids is only just beginning to be understood. Computer simulations
of thin hard plates with polydisperse diameters have shown, for
example, that the isotropic-nematic coexistence gap widens with
polydispersity~\cite{BatFre99}. (Though the usual caveat applies
regarding the results of semi-grandcanonical simulations, which
address the case of variable rather than fixed polydispersity.)  The
fractionation of plate {\em diameters} between I and N phases was
observed to be rather weak. Plates with polydisperse {\em
thicknesses}, on the other hand, displayed strong fractionation in
experiments on the I-N transition~\cite{VanVogLek00}.

At higher densities, experiments have shown the columnar phase to be
remarkably robust against polydispersity in plate
diameters~\cite{VanKasLek00}, tolerating polydispersities up to
$s\approx 25\%$. On further increasing the density, a crossover to
smectic ordering was observed; this seems plausible, since a spread in
particle diameters should prevent an efficient packing of the columns
of particles at high densities, favouring instead the layered structure of a
smectic.

The addition of non-adsorbing polymer produces further interesting
features in the phase behaviour of
polydisperse platelets. Experimentally, a strong
widening of the isotropic-nematic coexistence gap was
observed~\cite{VanVogLek00}, along with the occurrence of two separate
isotropic phases. The latter effect seems to be similar to the
``splitting'' of the hard sphere fluid into a gas and a liquid by
the addition of polymer.

Even more complex phase behaviour, finally, can occur in mixtures of
rod- and plate-like colloidal particles. Recent
experiments~\cite{VanLek00,VanLek00b} show dramatic polydispersity
effects: Up to five coexisting phases are found, rather than the
maximum of three expected for monodisperse hard rods and plates.

Most of the above results for systems involving plate-like colloids
remain poorly understood theoretically; open questions include, for
example, the contrasting effects of diameter and thickness
polydispersity at the I-N transition (especially as regards
fractionation), and the precise topologies of the phase diagrams for
plate-polymer and plate-rod mixtures. At least for the phenomena
involving isotropic and nematic phases, progress should be possible
using second
virial theories of the Onsager type; if the angular dependences are
truncated as described after eq.\eq{K_exp}, free energy expressions
with a truncatable structure will result and can be studied using for
example the moment free energy method.

\section{Outlook}
\label{sec:outlook}

Throughout this article, I have focussed entirely on equilibrium bulk
phase behaviour. Beyond this, there are significant open challenges in
understanding the effects of polydispersity on inhomogeneous phases
and on phase transition kinetics.

\subsection{Inhomogeneities}
\label{sec:inhomogen}

Inhomogeneities come to the fore when one is interested in, for
example, the behaviour of a polydisperse material near an interface or
a wall; one might want to calculate \eg\ the effect of
polydispersity on interfacial tensions or other interfacial
thermodynamic properties. The description of an inhomogeneous system requires
a particle density distribution $\rh(\rv,\sig)$ which
depends not only on the polydisperse attribute $\sigma$ but also 
on the spatial location $\rv$; the density distribution
$\rhosig$ used above to describe the state of bulk materials is found
from this by integration over the sample volume, $\rhosig = \int\!
d\rv\, \rh(\rv,\sig)$.  In dependence on $\rh(\rv,\sig)$ one can again
define a free energy $f([\rh(\rv,\sig)])$---conventionally referred to
as a ``density functional''---that assumes its minimal value at the
equilibrium density distribution $\rh(\rv,\sig)$ (see
\eg~\cite{Evans92,Lowen94} for reviews of density functional theory).

In principle this approach can also be used to obtain from first
principles the free energy of bulk phases with spatial ordering, such
as hard sphere crystals: To get the free energy $f([\rhosig])$ that I
have used throughout this paper, one would have to minimize
$f([\rh(\rv,\sig)])$ over all $\rh(\rv,\sig)$ with 
the given $\rhosig=\int\! d\rv\, \rh(\rv,\sig)$.
With orientational degrees of freedom included
appropriately, the same method would also apply \eg\ to the smectic,
columnar and crystalline phases of rod- and plate-like colloids. In
practice, this program can of course only be implemented very
approximately: To start with, the full free energy functional
$f[\rh(\rv,\sig)]$ is not known exactly; and the minimization over the
spatial density distribution can normally only be carried out over a
small number of assumed candidate structures, parameterized by
appropriate variational parameters.

What, then, are the specific challenges in the treatment of
inhomogeneities that arise from the presence of polydispersity?
Firstly, there is the problem of how to incorporate polydispersity
into the construction of approximate density functionals. For
polydisperse hard spheres, some significant progress in this direction
has been made recently by Pagonabarraga, Cates and
Ackland~\cite{PagCatAck00}, exploiting again a moment structure for the
excess part of the free energy: The moments $\ri$ then generalize to
spatially varying densities $\rho_i(\rv)$, defined as local averages
of the full density distribution $\rho(\rv,\sig)$. For spatially
extended objects such as polymers, the most appropriate way of
carrying out the local averaging is by no means
obvious~\cite{Warren99b}; a recent proposal models the polymers as
interpenetrable particles with a fixed monomer density profile about
their centre, chosen to reproduce the correct structure factor for
ideal polymer chains~\cite{PagCat01}.

The second challenge is to use density functionals for polydisperse
systems in practical calculations of interfacial properties etc. When
the excess free energy has a dependence only on certain spatially varying
moment densities $\rho_i(\rv)$, this can be done relatively
efficiently: The problem then effectively reduces to that of a
conventional density functional theory for a discrete mixture of
quasi-species~\cite{PagCatAck00,PagCat01}.

\subsection{Phase separation kinetics}

The kinetics of phase separation in polydisperse systems is a very
challenging and to a large extent unsolved problem. Above, we have
seen that description of equilibrium phase behaviour can be
substantially simplified through the use of \moms; a natural question
to ask is then whether \moms\ remain useful in understanding the
kinetics of phase separation.  Warren~\cite{Warren99b} has in fact
argued that in many systems the zeroth moment (total particle density)
$\rn$ should relax much more rapidly---by collective diffusion---than
can the higher moments, whose equilibration requires interdiffusion
of different particle species. This leads to the hypothesis that phase
separation could proceed in two stages: In the first stage, only the
densities of coexisting phases would equilibrate, while their
compositions would remain equal (``quenched'') to that of the parent
phase; the kinetics in the second stage would be much slower and bring
the compositions of the phases to equilibrium by fractionation.
Of course, the onset of
phase coexistence with all compositions quenched will generally occur
at a different point in the phase diagram than if full fractionation
is allowed. Experimentally observed cloud and shadow curves, for
example, could therefore be quite strongly dependent on the timescale
of a phase separation experiment, probing behaviour ranging from the
quenched to the fully fractionated phase diagram.

In principle, it is of course possible to treat the phase separation
kinetics in polydisperse systems by binning the range of the
polydisperse attribute $\sig$, reducing the problem to the dynamics of a
finite mixture of discrete species. In general one expects this
approach to be infeasible numerically; Clarke~\cite{Clarke01} has
recently shown, however, that it can be efficiently implemented to
study the early stages of phase separation of polydisperse polymers
which are suddenly cooled into a two-phase region of their phase
diagram.

Finally, there is the intriguing possibility that the kinetics of
phase separation (and, in particular, fractionation) in polydisperse
systems could be so slow as to make the equilibrium phase behaviour
unobservable in practice. Evans and Holmes~\cite{EvaHol01} have
recently argued that this is the case for polydisperse hard sphere
crystals: Once particles are incorporated into a crystal nucleus
growing from the hard sphere fluid, they essentially no longer diffuse
on experimental timescales. The size distribution of particles in the
crystal will thus be ``frozen in'', and determined by the mechanism of
crystal growth rather than the conditions of thermodynamic
equilibrium. A full understanding of such non-equilibrium effects on
the experimentally observed phase behaviour of colloidal systems
remains a significant challenge for future work.

\section{Conclusion}

In this article, I have attempted to give an overview of the current
state of the art in the field of polydisperse phase equilibria,
focussing on theoretical approaches for predicting coexistence
between bulk phases. Polydisperse systems are characterized by an
effectively infinite number of distinguishable particle species (and
thus of conserved densities), and this makes even the apparently
simple task of predicting phase equilibria from a known free energy
(functional) highly nontrivial. As reviewed in Sec.~\ref{sec:methods},
a number of methods have been developed to tackle this problem; the
most detailed understanding of phase behaviour can be achieved for
truncatable free energies, whose excess part depends only on a number
of moments of the density distribution $\rhosig$ rather than on all
its details. As shown in Sec.~\ref{sec:systems}, many (approximate)
free energies that can be used to describe polymeric and colloidal
system fall into this class. The phase behaviour that even these
relatively simple models produce is extremely rich compared to that of
monodisperse systems, and many intriguing questions remain unanswered.
The same is true, to an even greater degree, of the largely unexplored
areas of interfacial behaviour and phase separation kinetics which I
touched on briefly in Sec.~\ref{sec:outlook}

{\em Acknowledgements:} I have greatly benefited from stimulating and
helpful discussions with M E Cates, N Clarke, R M L Evans, T C B
McLeish, P Olmsted, I Pagonabarraga, W C K Poon, R Sear, A Speranza,
and P B Warren; I also acknowledge partial financial support from
EPSRC (GR/R52121/01).

\end{document}